\documentclass[iop,numberedappendix,apj,appendixfloats]{emulateapj}
\newcommand\ksmpc{{~km~s$^{-1}$~Mpc$^{-1}$}}

\newcommand\dell{$\Delta\log L$\xspace}
\newcommand\delt{$\Delta\log T$\xspace}

\newcommand\ebv{$\mathrm{E(B\!-\!V)}$\xspace}
\newcommand\btr{$(B/T)_r$\xspace}
\newcommand\btg{$(B/T)_g$\xspace}
\newcommand\zmin{$z_\mathrm{min}$\xspace}
\newcommand\zmax{$z_\mathrm{max}$\xspace}
\newcommand\gr{$g\!-\!r$\xspace}
\newcommand\pps{$P_\mathrm{pS}$\xspace}

\usepackage{xspace}
\usepackage{amsmath}
\usepackage{graphicx}
\usepackage[colorlinks,breaklinks, citecolor=blue,linkcolor=blue]{hyperref}
\usepackage{booktabs}
\usepackage{txfonts}
\usepackage{etoolbox}

\makeatletter
\patchcmd{\NAT@citex}
    {\@citea\NAT@hyper@{\NAT@nmfmt{\NAT@nm}\NAT@date}}
    {\@citea\NAT@nmfmt{\NAT@nm}\NAT@hyper@{\NAT@date}}
    {}
    {}
\patchcmd{\NAT@citex}
    {\@citea\NAT@hyper@{%
         \NAT@nmfmt{\NAT@nm}%
         \hyper@natlinkbreak{\NAT@aysep\NAT@spacechar}{\@citeb\@extra@b@citeb}%
         \NAT@date}}
    {\@citea\NAT@nmfmt{\NAT@nm}%
     \NAT@aysep\NAT@spacechar%
     \NAT@hyper@{\NAT@date}}
    {}
    {}
\patchcmd{\NAT@citex}
    {\@citea\NAT@hyper@{%
         \NAT@nmfmt{\NAT@nm}%
         \hyper@natlinkbreak
         {\NAT@spacechar\NAT@@open\if*#1*\else#1\NAT@spacechar\fi}%
         {\@citeb\@extra@b@citeb}%
         \NAT@date}}
    {\@citea\NAT@nmfmt{\NAT@nm}%
        \NAT@spacechar\NAT@@open\if*#1*\else#1\NAT@spacechar\fi%
        \NAT@hyper@{\NAT@date}}
    {}
    {}
\makeatother

\defcitealias{simard2011}{S11}
\defcitealias{lackner2012}{L12}
\defcitealias{maraston2005}{M05}
\defcitealias{bruzual2003}{BC03}

\slugcomment{Accepted on October 29, 2013 for Publication in Astrophysical Journal Supplements}

\shorttitle{A catalog of stellar masses for the SDSS DR7}
\shortauthors{Mendel et al.}

\begin{document}

\title{A catalog of bulge, disk, and total stellar mass estimates for the Sloan Digital Sky Survey}

\author{J.\ Trevor Mendel\altaffilmark{1,2,$\dagger$}, Luc Simard\altaffilmark{3}, Michael Palmer\altaffilmark{2}, Sara L. Ellison\altaffilmark{2}, \and David R. Patton\altaffilmark{4}}

\altaffiltext{$\dagger$}{\texttt{jtmendel@mpe.mpg.de}}
\altaffiltext{1}{Max-Planck-Institut f\"{u}r Extraterrestrische Physik, Gie\ss enbachstra\ss e, 85748 Garching, Germany}
\altaffiltext{2}{Department of Physics and Astronomy, University of Victoria, Victoria, British Columbia, V8P 1A1, Canada}
\altaffiltext{3}{National Research Council of Canada, 5071 West Saanich Road, Victoria, British Columbia, V9E 2E7, Canada}
\altaffiltext{4}{Department of Physics \& Astronomy, Trent University, 1600 West Bank Drive, Peterborough, Ontario, K9J 7B8, Canada}

\begin{abstract}

We present a catalog of bulge, disk, and total stellar mass estimates for $\sim$660,000 galaxies in the Legacy area of the Sloan Digital Sky Survey Data Release 7.  These masses are based on a homogeneous catalog of $g$- and $r$-band photometry described by \citet{simard2011}, which we extend here with bulge+disk and S\'ersic profile photometric decompositions in the SDSS $u$, $i$, and $z$ bands.  We discuss the methodology used to derive stellar masses from these data via fitting to broadband spectral energy distributions (SEDs), and show that the typical statistical uncertainty on total, bulge, and disk stellar mass is $\sim$0.15 dex.  Despite relatively small formal uncertainties, we argue that SED modeling assumptions, including the choice of synthesis model, extinction law, initial mass function, and details of stellar evolution likely contribute an additional 60\% systematic uncertainty in any mass estimate based on broadband SED fitting.  We discuss several approaches for identifying genuine bulge+disk systems based on both their statistical likelihood and an analysis of their one-dimensional surface-brightness profiles, and include these metrics in the catalogs.  Estimates of the total, bulge and disk stellar masses for both normal and dust-free models and their uncertainties are made publicly available here.

\end{abstract}

\keywords{galaxies: fundamental parameters -- galaxies: statistics -- galaxies: bulges}

\section{Introduction}
\label{intro}

Galaxies' observable properties are closely tied to their stellar mass, either directly, or indirectly through correlations with their host environment (e.g. dark matter halo mass).  More massive galaxies are found to be on average redder \citep[e.g][]{blanton2003a,baldry2006}, older \citep{proctor2002,kauffmann2003a,gallazzi2005}, more metal-rich \citep{tremonti2004,gallazzi2005}, more strongly clustered \citep[e.g.][]{beisbart2000,skibba2006}, and more likely to host active galactic nuclei \citep[e.g.][]{kauffmann2003} than their less-massive counterparts.  Measuring a given galaxy's stellar mass therefore provides a snapshot of its evolutionary state.  Unfortunately, this apparent simplicity is betrayed by the complex combination of gas accretion, merging, and star-formation events that govern the buildup of stellar mass over cosmic time.  Quantifying the relative contribution of these processes to galaxies' evolution remains an as-yet unsolved problem.

It has been known for some time that the bimodal relationship between galaxies' color and stellar mass \citep{strateva2001,baldry2004} is also reflected in the details of their physical structure: the population of blue galaxies is dominated by rotationally-supported disks, while red galaxies are predominantly classified as spheroidal \citep[e.g.][]{lintott2008}.  \citet{driver2006} suggest that the relationship between galaxy color (or, alternatively, star formation) and structure can be explained by the differing properties of bulges and disks.  The observed \emph{galaxy} bimodality is then a direct result of mixing these two different components throughout the galaxy population.  From a theoretical standpoint, these distinct structural components arise from different formation mechanisms: disks form as a result of angular momentum conservation when gas is accreted and cools in dark matter halos \citep{white1978,fall1980}, while spheroids are thought to form as a result of merger events \citep{toomre1972,negroponte1983} and internal (secular) processes \citep[e.g.][]{kormendy1982a}.  Analyzing the structural properties of bulges and disks therefore holds significant astrophysical interest, insomuch as they encode the details of galaxies' evolution.

Thanks to the ever-increasing scale of galaxy surveys and improvements in computing capabilities, it has become possible to routinely analyze the morphological properties of tens to hundreds of thousands of galaxies \citep[e.g.][]{allen2006,benson2007,simard2011,lackner2012,barden2012,kelvin2012}.  These large datasets provide a framework within which to study the relative properties of galaxies as well as their structural sub-components, and the means to undertake a statistical comparison between observations and theory.  Recently, \citet[hereafter \citetalias{simard2011}]{simard2011} published a catalog of revised $g$- and $r$-band photometry for $\sim$1.12 million galaxies in the Sloan Digital Sky Survey (SDSS) which acts as the jumping-off point for the determination of stellar masses discussed here.  The \citetalias{simard2011} catalog contains two important updates relative to the standard SDSS photometry that motivate the present work.  First, in addition to measuring galaxies' total flux, \citetalias{simard2011} also model galaxies' two-dimensional surface-brightness profiles as the sum of a de Vaucouleurs bulge and exponential disk.  For each galaxy we therefore have an estimate of its bulge and disk size, position angle, inclination and ellipticity, as well as the global bulge-to-total flux ratio.  Second, \citetalias{simard2011} re-estimate the local sky background and segmentation mask on a galaxy-by-galaxy basis, providing improved photometric estimates for galaxies in crowded environments, and in particular close pairs \citep[e.g.][]{patton2011}.

In this paper we utilize an expanded version of the \citetalias{simard2011} photometric catalog which includes $u$-, $i$-, and $z$-band measurements to estimate bulge, disk, and total stellar masses for $\sim$660,000 galaxies from the SDSS Legacy Survey.  Our goal in this work is to describe the methodology used to derive stellar masses from broadband photometry, as well as provide an assessment of the uncertainties associated with this process that should be borne in mind with their use.  Together with the structural parameters presented by \citetalias{simard2011}, this catalog represents the largest currently-available database of bulge and disk properties in the local Universe. The paper is organized as follows.  In Section \ref{data} we describe the sample of SDSS galaxies and their corresponding photometric data on which our stellar mass estimates are based.  In Section \ref{mass_estimates} we discuss the construction of our Stellar Population Synthesis (SPS) grid and outline the methods used to determine stellar masses and their uncertainties.  In Section \ref{uncertainties} we provide a brief discussion of the potential systematic uncertainties on our mass estimates, including the choice of SPS model, initial mass function, extinction law, and details of stellar evolution.  Finally, in Section \ref{discussion} we discuss details of the catalog, and provide guidelines for identifying subsamples of bulges and disks for subsequent study.  For users primarily interested in understanding the overall reliability of the stellar masses presented here we recommend skipping ahead to Section \ref{uncertainties}.

Throughout this paper we adopt a flat cosmology with $\Omega_{\Lambda} = 0.7$, $\Omega_{\mathrm{M}} = 0.3$ and $H_0 = 70$\ksmpc.  All magnitudes are given in the AB system, and we quote luminosities in units of $L_\odot$, adopting AB magnitudes of the sun from \citet{blanton2007a}.

\section{Photometric Data}
\label{data}

\citetalias{simard2011} present a catalog of photometric measurements for the SDSS legacy survey Data Release 7 \citep[DR7;][]{abazajian2009} based on a reprocessing of the SDSS images, which we extend here with the addition of the $u$, $i$, and $z$ bands.  This revised set of photometry forms the basis for our derivation of galaxies' bulge, disk, and total stellar masses.  We review the salient features of these data below, and refer the reader to \citetalias{simard2011} for details of the reprocessing procedure.

The underlying data of \citetalias{simard2011}'s photometry are the bias-subtracted, flat-fielded images output by the SDSS {\sc frames} pipeline.  Individual objects were identified in these images using {\sc sextractor} \citep{bertin1996}, and the resulting segmentation image was used to identify and mask out neighboring objects in the subsequent photometric measurements.  Local sky levels were then re-determined on an object-by-object basis using a minimum of 20,000 sky pixels, excluding pixels within $4\farcs0$ of any primary or neighboring object mask. 

Photometric measurements were obtained from parametric fits to galaxies' two-dimensional surface-brightness distributions using the {\sc gim2d} software package \citep{simard2002}, taking into account the object- and band-specific point spread function (PSF).  Fits to the $u$, $g$, $i$, and $z$ bands were each performed simultaneously with the $r$ band by forcing structural parameters (e.g. bulge and disk size, position angle, S\'ersic index, etc.) to be the same in both bandpasses.  Remaining parameters---total flux, bulge-to-total flux ratio ($B/T$), and image registration relative to the model centroid---were fit independently in each bandpass.  In the case of the $u$-, $i$-, and $z$-band fits we fix the galaxy structural parameters (with the exception of S\'ersic index; see Appendix \ref{fit_comp}) to the values derived from the combined \emph{gr} fits published by \citetalias{simard2011}.  This approach provides a robust estimate of galaxies' multi-band SED, albeit at the expense of flexibility offered by more sophisticated wavelength-dependent models \citep{hausler2013,vika2013}.

In total, \citetalias{simard2011} provide structural decompositions for 1,123,718 objects with extinction-corrected $r$-band Petrosian magnitudes $14 < m_r < 18$.  The bright-end magnitude limit is adopted to avoid problematic deblending of luminous galaxies in the SDSS catalogs.  Here we further restrict ourselves to those objects which are spectroscopically classified as galaxies ({\tt specClass = 2}) and satisfy the SDSS main galaxy sample selection criteria \citep[$m_r \leq 17.77$;][]{strauss2002} over a redshift range of $0.005\leq z \leq 0.4$.  These criteria identify a subsample of 669,634 galaxies which are then passed to {\sc gim2d}.  Profile fits failed to converge in one or more bands for 4,648 objects ($\sim$0.7\% of the sample), and these are necessarily removed from the sample.  We exclude an additional 5,614 galaxies whose {\sc gim2d}-derived photometry is contaminated by nearby bright stars or in poor agreement with the SDSS fiber aperture magnitudes\footnote{Specifically, we require that the $g-r$ color in a $3^{\prime\prime}$ aperture derived from the best-fit, PSF-convolved {\sc gim2d} model is within $\pm$0.2 mag of the fiber color reported by the SDSS.} (see also the discussion by \citetalias{simard2011}).  Our final sample contains 659,372 galaxies with updated $ugriz$ photometry.  These selection criteria are summarized in Table \ref{sample_numbers}.

\citetalias{simard2011} describe three different models used in their fitting procedure: a single-component S\'ersic profile, a two-component de Vaucouleurs bulge plus exponential disk, and a two-component S\'ersic bulge plus exponential disk.  In what follows we focus on the modelling of galaxies as either a single-component S\'ersic profile or a de Vaucouleurs bulge plus exponential disk, as the SDSS data are generally insufficient to provide good constraints on bulge light profiles when the shape parameter $n$ is left unconstrained \citepalias[see also \citealp{lackner2012}]{simard2011}.  The systematic uncertainties on total flux recovered by \citetalias{simard2011}'s bulge+disk and S\'ersic decompositions have been estimated using simulated galaxy data, and are $\lesssim$0.15 mag in most cases depending on the assumed "true" galaxy surface brightness profile.  Individual bulge and disk fluxes are accurate to within $\sim$0.3 mag in most cases, but become poorly constrained at extreme values of $B/T$.  See Appendix \ref{flux_sims} for a more in depth discussion of photometric uncertainties in the \citetalias{simard2011} catalogs.

\begin{deluxetable}{lr}
\tablewidth{0pc}
\tablecaption{Summary of sample selection criteria}
\tablehead{\colhead{Selection criterion} & \colhead{Galaxies remaining (removed)}}
\startdata
\citet{simard2011} sample & 1,123,718 \\
Spectroscopic sample & 669,634 (454,084) \\
Successful \emph{ugriz} decomposition & 664,986 (4,648) \\
Quality control & 659,372 (5,614)
\enddata
\tablecomments{Each line shows the remaining galaxies (number of galaxies removed) after applying each of our different selection criteria.  See Section \ref{data} for more details. \label{sample_numbers}} 
\end{deluxetable}

\section{Stellar mass estimates}
\label{mass_estimates}

Our general approach to estimating stellar mass involves the comparison of each galaxy's observed SED to a library of synthetic stellar populations.  This library is constructed so that models span a range of ages, metallicities, star-formation histories, and dust properties observed in nearby galaxies.  The relevant parameters of our SPS grid are summarized in Table \ref{grid_param}.  The basic mechanics of this SED fitting procedure have been outlined by numerous authors \citep[e.g.][]{brinchmann2000,kauffmann2003b,bundy2006,salim2007,taylor2011,swindle2011,sawicki2012}, and are described in detail below.

\subsection{Stellar population synthesis}
\label{models}

\begin{figure}
\centering
\includegraphics[scale=0.90]{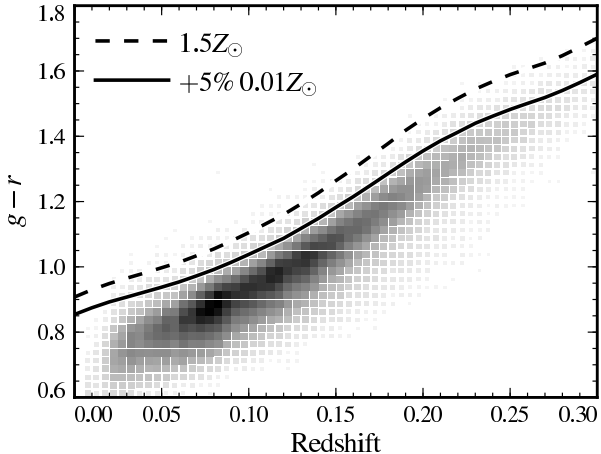}
\caption{Evolution of \gr color as a function of redshift for red sequence galaxies (shading).  Lines show the predicted color evolution for two different stellar populations, one with a single $Z=1.5Z_\odot$ population (dashed line) and another that includes 5\% by mass in metal poor stars ($Z=0.01Z_\odot$; solid line).  Both populations have $\tau=0.1$~Gyr, and are normalized to an age of 12.0~Gyr at $z=0$.  The inclusion of a metal-poor subcomponent provides a much better description for the upper envelope of the observed galaxy population, and is adopted as our fiducial model for estimates of stellar mass (see Section \ref{models} for details).}
\label{fig:color_evo}
\end{figure}

The basis for our SPS model grid is the Flexible Stellar Population Synthesis (FSPS) code of \citet{conroy2009}\footnote{available at \url{http://www.ucolick.org/~cconroy/FSPS.html}}, which allows us to generate synthetic SEDs given a range of galaxy properties.  All models are constructed using Padova isochrones \citep{girardi2000,marigo2007,marigo2008} and the BaSeL3.1 theoretical stellar library \citep{lejeune1997,lejeune1998,westera2002}.  We adopt a form for the stellar initial mass function (IMF) from \citet{chabrier2003} with lower and upper integration limits of 0.08 and 120 M$_\odot$.  We refer the reader to \citet{conroy2009} for details of the FSPS model construction, as well as subsequent papers \citep{conroy2010a,conroy2010b} that describe the calibration of FSPS models with observational data.

We generate a suite of models with smoothly-declining star-formation histories characterized by an $e$-folding timescale, $\tau$, such that the time-dependent star-formation rate $\psi_\star(t)$ is given by $\psi_\star\left(t\right) \propto \tau^{-1} \exp\left({-t/\tau}\right)$.  We follow \citet{taylor2011} in sampling $\tau$ on a semi-uniform grid where $\log(\tau/\mathrm{yr})$ spans the range 8--9 in steps of 0.2 dex, and from 9.1--10 in steps of 0.1 dex.  Stellar metallicities cover the range $-1.8\leq\log\left(Z/Z_\odot\right) \leq 0.2$ in steps of 0.2 dex.  We include internal attenuation by dust using the extinction law of \citet{calzetti2000} with \ebv sampled uniformly over the range $0 \leq \mathrm{E(B\!-\!V)} \leq 1$ (i.e. a range of $V$-band optical depth of  $0 \leq \tau_V \leq 2.855$) in steps of 0.05 mag.  Finally, population ages are sampled uniformly with $8 \leq \log\left(t/\mathrm{yr}\right) \leq 10.1$ in steps of 0.05 dex.  The full SPS grid therefore contains 158,928 individual sets of synthetic $ugriz$ photometry.  These properties are summarized in Table \ref{grid_param}.  We also perform a set of fits where we remove dust as a free parameter and instead fix \ebv = 0.  We discuss a comparison of dusty and dust-free models in Section \ref{dust_free}.

There is a well-documented discrepancy between the observed colors of red galaxies and the predicted colors of old, metal rich populations in SPS models such that synthetic \gr photometry is $\sim$0.1 mag redder than is observed \citep[e.g.][]{eisenstein2001,maraston2009,conroy2010a}.  This is illustrated in Figure \ref{fig:color_evo}, where we plot apparent \gr color as a function of redshift for red galaxies in the SDSS.  When compared to the predicted color evolution of a dust-free, metal-rich single stellar population (dashed line in Figure \ref{fig:color_evo}), there is a clear tendency for the models to be redder than observations by 0.05 to 0.1 mag.

\begin{deluxetable}{lcc}
\tablewidth{0pc}
\tablecaption{SPS Grid Parameters}
\tablehead{\colhead{Parameter} & \colhead{Description} & \colhead{Range of values}}
\startdata
$\tau$	& $e$-folding time	& $8 \leq \log(\tau/\mathrm{yr}) \leq 10$ in 16 steps \\ 
$Z$		& stellar metallicity	&  $-1.8\leq\log (Z/Z_\odot) \leq 0.2$ in 11 steps\\ 
$\mathrm{E(B\!-\!V)}$	& color excess	& $0 \leq \mathrm{E(B\!-\!V)} \leq 1$ in 21 steps  \\ 
$t$	& population age & $8 \leq \log (t/\mathrm{yr}) \leq 10.1$ in 43 steps \\
IMF & stellar IMF & Chabrier 2003 \\
$k(\lambda)$ & extinction law & Calzetti et al. 2000 
\enddata
\tablecomments{Metallicities are computed adopting $Z_\odot = 0.019$.  See Section \ref{models} for details of the SPS grid construction. \label{grid_param}} 
\end{deluxetable}

\begin{figure*}
\centering
\includegraphics[scale=0.90]{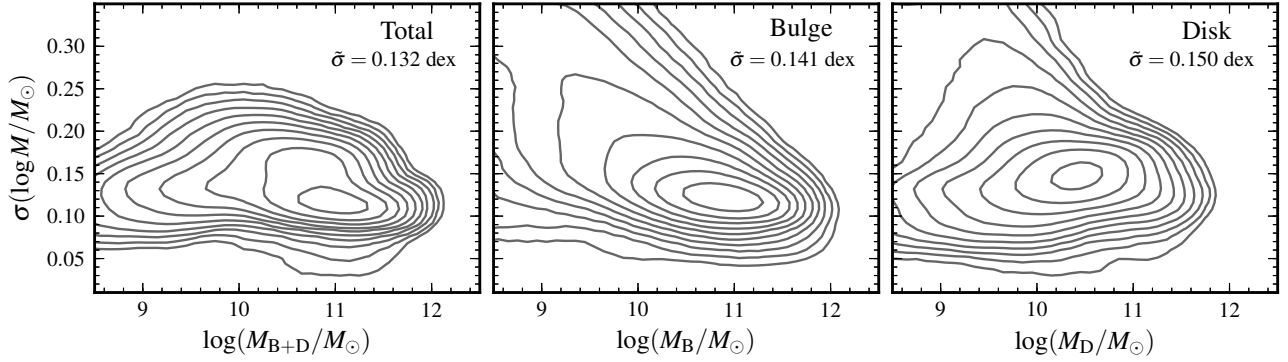}
\caption{Contours of uncertainty versus stellar mass for total, bulge and disk components.  In each panel we quote the median (statistical) uncertainty on the derived stellar mass.}
\label{fig:mass_error}
\end{figure*}

Barring changes in our general picture of stellar evolution, rectifying the predicted colors from SPS models with observational data requires including a minority population of hot, blue stars.  These can come in a variety of forms, either as a distinct evolutionary phase (e.g. blue straggler and/or blue horizontal branch stars) or as a more complex stellar population (e.g an $\alpha$-rich or metal-poor subcomponent).  As an example, the solid line in Figure \ref{fig:color_evo} shows the predicted color evolution of a stellar population that includes 5\% by mass in metal-poor stars with $\log (Z/Z_\odot) = -1.98$.  The net effect of including this metal-poor sub-component is relatively minor in terms of the derived mass-to-light ratio (\emph{M/L})---masses are on average 0.02 dex lighter relative to a simple single-metallicity population; however, it improves the overall agreement between observed and synthetic $ugriz$ SEDs \citep[see also the discussion by][]{conroy2010b}.  While arguments following chemical evolution support the incorporation of metal-poor stars into SPS models \citep[e.g.][]{pagel1975} they are not the only solution to the problem highlighted by Figure \ref{fig:color_evo}.  Instead it is likely that a combination of variables contribute to the observed offset between SPS models and data, and it is beyond the scope of the present work to investigate this issue in more detail.  Nevertheless, given the improved agreement between the observed and synthetic SEDs when incorporating even a small component metal-poor stars we choose to include such a subpopulation in our fiducial SPS model.  We stress that the exact properties of the included metal-poor sub-component (in terms of mass fraction and metallicity) are not unique.

Finally, in contrast to the widely-used stellar mass catalogs produced by the MPA-JHU group \citep{kauffmann2003b,salim2007}, our SPS grid does not include bursty star-formation histories.  With the exception of recent starbursts, the majority of local galaxies' SEDs are well fit by single component $\tau$ models due to the smooth evolution of SED properties with population age.  More importantly, properly accounting for bursty star-formation histories is non-trivial.  Such galaxies are not easily identified using photometry alone, and the ability to recover reliable \emph{M/L} estimates depends on the relative frequency with which bursts are incorporated into the underlying SPS library.  \citet{gallazzi2009} show that including too large a fraction of bursty SPS models can lead to a systematic underestimate of \emph{M/L} by up to 0.1 dex for galaxies whose star-formation histories are generally smooth, while fitting bursty galaxies with smooth models may overestimate their \emph{M/L} by up to 0.2 dex.  Our decision to exclude bursts is therefore predicated on obtaining robust mass estimates for the \emph{majority} of (smoothly evolving) galaxies.  Conversely, for the minority population of bursty galaxies \citep[$<$10\%;][]{kauffmann2003b}, we may overestimate stellar masses by a factor of 2 or more, and systematically underestimate their uncertainties \citep[see also][]{walcher2011}.

\subsection{SED fitting}

For each photometric object---galaxy, disk, or bulge---our goal is to determine the set of fundamental parameters that most accurately describe its underlying stellar population.  It is through this determination that we ultimately derive estimates of stellar mass.  We have chosen to approach this task via Bayes' Theorem, which allows us to compute the conditional probability for a set of model parameters, $\theta$, given the observational data as 

 \begin{equation}
 p(\theta|F_\mathrm{obs}) \propto p(F_\mathrm{obs}|\theta) p(\theta).
 \label{bayes}
 \end{equation}

\noindent In the above, $p(F_\mathrm{obs}|\theta)$ is the likelihood of observing a set of fluxes, $F_\mathrm{obs}$, given $\theta$, and $p(\theta)$ is the prior probability of observing $\theta$ before considering the observations.  In general $\theta$ specifies a vector, in this case the defining parameters of our SPS grid---i.e. $\tau$, $Z$, \ebv, and $t$.  One benefit of adopting a Bayesian approach to SED fitting is that our output, $p(\theta|F_\mathrm{obs})$, reflects the full posterior probability density function (PDF), incorporating uncertainties on individual parameters of the SPS grid in our final estimates of stellar mass.

In the case where an object is detected in all bands, the likelihood function for model $i$ is defined as $p(F_\mathrm{obs}|\theta_i) \propto \exp(-\chi_i^2/2)$, where $\chi_i^2$ is given by

\begin{equation}
\chi_i^2  = \sum_X \left(\frac{F_{\mathrm{obs},X} - A_i F_{\mathrm{mod}_i,X}}{\sigma\left(F_{\mathrm{obs},X}\right)}\right)^2.
\label{chi2}
\end{equation}

\noindent Here the summation is over the photometric passbands, i.e. $X = (u, g, r, i, z)$.  $F_{\mathrm{obs},X}$ and $F_{\mathrm{mod}_i,X}$ are the observed and model fluxes through passband $X$ respectively, and $\sigma(F_{\mathrm{obs},X})$ is the uncertainty on $F_{\mathrm{obs},X}$.  The normalization factor $A_i$ accounts for the scaling between the model SED (which is normalized to a \emph{time-integrated} mass of 1 M$_\odot$) and the observational data, and can be computed analytically by taking the derivative of Equation \ref{chi2} with respect to $A_i$.

While galaxies are always detected in all 5 bands, in some cases bulges and disks may be undetected in one or more photometric passbands, for example when $B/T$ = 0 or 1.  In order to incorporate information from these non-detections in our SED fits we modify Equation \ref{chi2} following \citet{sawicki2012} such that 

\begin{align}
\chi_i^2  &= \sum_X \left(\frac{F_{\mathrm{obs},X} - A_i F_{\mathrm{mod}_i,X}}{\sigma\left(F_{\mathrm{obs},X}\right)}\right)^2 \nonumber \\
&\quad {} -2\sum_Y \ln \int_{-\infty}^{F_\mathrm{lim,Y}} \exp\left[-\frac{1}{2}\left(\frac{F - A_i F_{\mathrm{mod}_i,Y}}{\sigma\left(F_{\mathrm{obs},Y}\right)}\right)^2\right] dF,
\label{mod_chi2}
\end{align}

\noindent where the first term is now the sum over \emph{detected} passbands and the second term accounts for the probability of a non-detection given the band-specific flux limit, $F_\mathrm{lim,Y}$, and model flux.  Note that in the case where an object is detected in all passbands Equation \ref{mod_chi2} reduces to the standard $\chi^2$ form of Equation \ref{chi2}.  There is no simple analytic solution for $A_i$ when incorporating non-detections into the likelihood function, so we instead solve for the normalization numerically.  

\begin{figure*}
\centering
\includegraphics[scale=0.95]{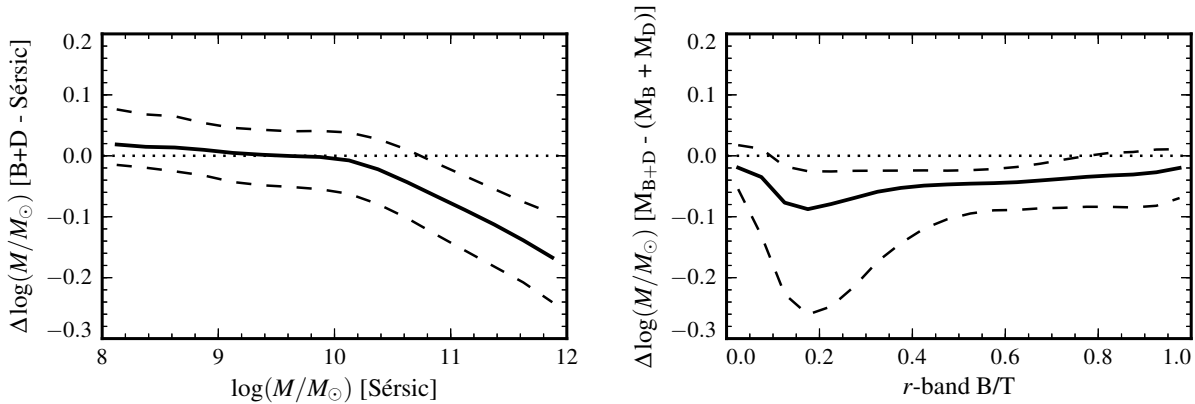}
\caption{Comparison of stellar mass derived from the bulge+disk and S\'ersic photometric catalogs described in Section \ref{data}.  In the left panel we plot the comparison of total masses derived from the bulge+disk and S\'ersic fits.  The solid line shows the median offset, while the dashed lines show $\pm$1 sigma percentiles.  In the right panel we show a comparisons of the total mass derived from the bulge+disk fits, $M_\mathrm{B+D}$, and the sum of the (independently determined) bulge and disk masses, $M_\mathrm{B} + M_\mathrm{D}$.  Solid and dashed lines again show the median and $\pm$1 sigma percentiles of the distribution.}
\label{fig:mass_comp}
\end{figure*}

\subsubsection{Priors}

There are two sets of prior assumptions that enter in to our SED fitting methodology.  The first are implicit, and contribute through choices made in constructing our SPS model grid.  These priors encompass our assumptions regarding the stellar initial mass function and extinction law, the range of observable ages, metallicities, and dust attenuations, as well as our assumption that the majority of galaxies are well described by smooth, exponentially declining star-formation histories.  

The second set of priors reflect our assumptions about the distribution of $\theta$---that is, how ``true'' stellar populations are likely to populate the parameter space defined by $t$, $\tau$, $Z$, and \ebv---and must be explicitly defined before we can compute the posterior PDF.  Here we choose to remain relatively agnostic to the distribution of stellar parameters by adopting priors that are uniform in $t$, $\log\tau$, $\log (Z/Z_\odot)$, and \ebv.

In some cases the photometric data are insufficient to constraint the properties of the underlying stellar population, particularly \ebv, when fitting for the bulge and disk stellar masses.  We therefore adopt the posterior PDF for \ebv from fits to the total photometry as a prior on \ebv in the bulge and disk fits.  In most cases the differences between using this empirical prior for the bulge and disk fits results in only minor changes relative to a uniform dust prior.

\subsection{Deriving stellar mass}

For each set of $F_\mathrm{obs}$ and $\theta_i$ the most likely stellar mass is given by $A_i M_{\mathrm{mod}_i, \star}$ with probability $p(\theta_i|F_\mathrm{obs})$\footnote{Subject to the constraint that $\int p(\theta|F_\mathrm{obs})d\theta \equiv 1$, i.e. that our model grid encompasses the full parameter space of galaxy properties.}, where $M_{\mathrm{mod}_i, \star}$ is the normalized stellar mass for model $i$.  For each galaxy or structural component, the stellar mass posterior PDF is obtained through the application of Equation \ref{chi2} or \ref{mod_chi2} at each point in our synthetic photometric grid, and is completely specified for a given set of $ugriz$ fluxes and their associated uncertainties.  

Uncertainties on the observed fluxes reflect a combination of the statistical uncertainties on the model parameters, output by {\sc gim2d}, and the r.m.s.~noise in the sky background, which we estimate using the statistics in each SDSS imaging field.  We include an additional uncertainty of 3\% in $griz$ and 5\% in $u$ to account for both the relative and absolute SDSS photometric calibration.  In most cases the photometric calibration and sky background both contribute significantly to the overall uncertainty, with the exception of the $u$ and $z$ bands where the background dominates.  In Appendix \ref{flux_sims} we present an analysis of the uncertainties on the \citetalias{simard2011} photometry based on Monte Carlo simulations.  We have verified that uncertainties estimated based on the {\sc gim2d} and image field statistics (described above) are consistent with those derived from the model galaxies presented in Appendix \ref{flux_sims}; we preferentially use the former as they are independent of the specific form of the galaxy surface brightness profile (i.e. S\'ersic or bulge+disk) used to generate the synthetic galaxies described in Appendix \ref{flux_sims}.

Final estimates of stellar mass are taken as the median of the $AM_{\mathrm{mod}, \star}$ distribution, weighted by $p(\theta|F_\mathrm{obs})$.  We account for foreground (Galactic) extinction using the \citet{schlegel1998} dust maps, and apply this correction directly to the FSPS photometry on a galaxy-by-galaxy basis.  We have not corrected for the contribution of emission lines to the observed broadband flux measurements, however for the majority of galaxies their contribution is negligible.  We quote uncertainties as the $\pm$1 sigma percentiles of the marginalized posterior PDF.  Note that we fit the total, bulge, and disk photometry independently, and therefore it is not necessarily true that the total, bulge, and disk masses---$M_\mathrm{B+D}$, $M_\mathrm{B}$, and $M_\mathrm{D}$, respectively---are consistent (i.e., we don't explicitly impose that $M_\mathrm{B} + M_\mathrm{D} = M_\mathrm{B+D}$).  Nevertheless, in 91\% of galaxies $M_\mathrm{B+D}$ and $M_\mathrm{B} + M_\mathrm{D}$ differ by less than their 1$\sigma$ uncertainties.  This comparison will be discussed further in Section \ref{mass_comp}.

The median (statistical) uncertainties of the total, bulge and disk stellar masses are $\sim$0.13, 0.14 and 0.15 dex, respectively.  Uncertainties on the total mass derived from bulge+disk or S\'ersic photometric models are similar.  In Figure \ref{fig:mass_error} we show the distribution for these three components as a function of derived stellar mass to give a sense of their behavior.

\subsection{S\'ersic vs. bulge+disk models}
\label{mass_comp}

In Figure \ref{fig:mass_comp} we show a comparison of masses derived from the bulge+disk and S\'ersic photometry.  In the left hand panel, we plot the difference in total mass from the bulge+disk and S\'ersic fits as a function of stellar mass derived from the S\'ersic photometry.  At masses below $\log (M/M_\sun) = 10$ the two estimates are remarkably similar, with a median offset of 0.02 dex.  On the other hand, at $\log (M/M_\sun) > 10$ the masses derived from S\'ersic photometry are generally larger than those from the bulge+disk fits by up to 0.2 dex.  This is consistent with the generally higher luminosities measured using S\'ersic profiles relative to other parametric or curve-of-growth approaches \citep[see, e.g.][]{bernardi2012,bernardi2013}.  The simulations discussed in Appendix \ref{flux_sims} suggest that the tendency for S\'ersic profiles to recover higher luminosities is actually a combination of two different effects: the known correlation between S\'ersic index and stellar mass \citep[e.g.][]{graham2003}, and the tendency for our bulge+disk models to systematically underestimate the luminosity of galaxies with $n > 5$ (Appendix \ref{flux_sims}). 

In the right hand panel of Figure \ref{fig:mass_comp} we plot the difference between the mass derived from the total bulge+disk photometry, $M_\mathrm{B+D}$, and the sum of the independently derived bulge and disk masses, $M_\mathrm{B} + M_\mathrm{D}$, as a function of the $r$-band bulge fraction, \btr.  This provides an indicator of the \emph{internal} consistency of the bulge+disk stellar masses.  At $(B/T)_r \gtrsim 0.4$ the sum of the bulge and disk masses typically agree to within 10\% of the mass derived from the total galaxy photometry, where $M_\mathrm{B} + M_\mathrm{D}$ is generally heavier.  This systematic offset can be traced back to the larger photometric uncertainties for the individual structural components.  

The comparison at low \btr is somewhat more unsettling: while the median offset stays relatively unchanged, in some cases $M_\mathrm{B} + M_\mathrm{D}$ exceeds $M_\mathrm{B+D}$ by more than a factor of 2.  The behavior of $M_\mathrm{B} + M_\mathrm{D}$ at low \btr is driven by $\sim$10\% of the total sample where the bulge masses are significantly overestimated.  Many of these ``problem'' bulges have extremely red colors (their median $g-r$ color is $\sim$1.4) and, in some cases, relatively large photometric uncertainties.  As a result, the posterior PDFs for these galaxies are dominated by families of heavily extincted models despite our use of a prior on \ebv for the bulge and disk fits.  The resulting bulge mass fractions for these systems are physically unreasonable given their typical photometric $B/T$ in the $i$ and $z$ bands of $\sim$25\%, and they should be approached with extreme caution.  In general, the comparison of the $M_\mathrm{B+D}$ and $M_\mathrm{B} + M_\mathrm{D}$ can be used as an indicator of the internal consistency of the bulge and disk photometry and mass estimates, and provides a useful diagnostic for identifying robust subsamples of the data.  We will discuss this point further in Section \ref{discussion}.

\subsection{Dusty vs. dust-free models}
\label{dust_free}

\begin{figure}
\centering
\includegraphics[scale=0.80]{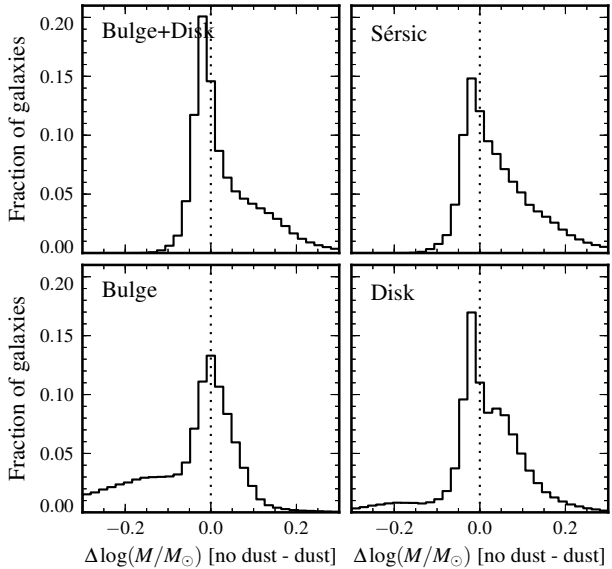}
\caption{Comparison of stellar mass derived from the default ``dusty'' fits and the dust-free fits described in Section \ref{dust_free}.  We show the offset in mass for each of the separate sets of photometry: bulge+disk (top left), S\'ersic (top right), bulge (bottom left), and disk (bottom right).  Vertical dotted lines indicate no offset.}
\label{fig:dust_comp}
\end{figure}

\begin{figure}
\centering
\includegraphics[scale=0.95]{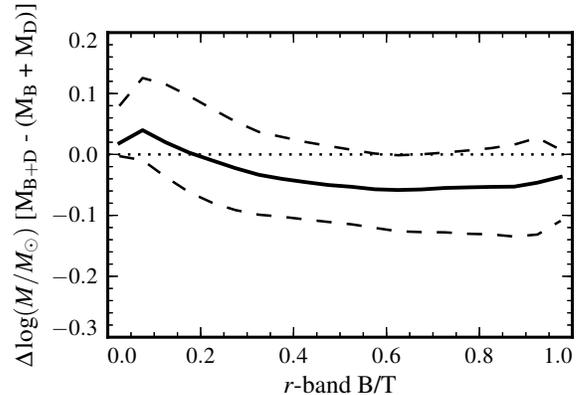}
\caption{Comparison of $M_\mathrm{B+D}$ and $M_\mathrm{B} + M_\mathrm{D}$ for the dust-free fits describe in Section \ref{dust_free}.  The solid and dashed lines show the median offset and $\pm$1 sigma percentiles.}
\label{fig:bd_nodust}
\end{figure}

Noisy or spurious photometric data can lead to unpredictable results when fitting broadband SEDs, and the combination of significant photometric uncertainties and degeneracy between dust extinction and stellar age can result in anomalously high stellar mass estimates \citep[e.g.][]{pforr2012}.  In the particular case of extremely red objects, such as those discussed in the previous Section,  masses can be significantly overestimated when using the full SPS grid due to the abundance of extincted models.  Therefore, we have produced an additional set of bulge+disk and S\'ersic stellar mass estimates in which \ebv has been set to zero.  These ``dust-free'' models are useful in instances where very uncertain and/or bad photometry result in spurious estimates of the stellar mass, particularly for bulges and disks.

In Figure \ref{fig:dust_comp} we show a comparison of the dusty and dust-free models for each of the photometric components.  In each case we plot the mass derived from the dust-free model minus the mass derived based on the full SPS grid.  In terms of the total photometry, fixing \ebv = 0 results in a relatively minor systematic shift in the stellar masses of $\sim$0.04 dex when using either the bulge+disk or S\'ersic photometry.  In both instances there is a tail of galaxies whose masses are up to 0.2 dex higher in the dust-free fits.  This tail is populated primarily by relatively red star-forming galaxies---that is, they populate the red side of the blue cloud in the color vs. stellar mass plane.  In general terms, removing dust as a free parameter makes blue galaxies heavier by $\sim$0.1 dex, while red galaxies get lighter by $\sim$0.03 dex.

Turning to the individual sub-components, we find no evidence for a systematic offset in bulge stellar mass when dust is excluded for most galaxies; this confirms our expectation that spheroids contain relatively little dust.  However, there is a tail of the bulge population whose masses are significantly higher when dust is included as a free parameter.  These are the particularly red, low $B/T$ systems highlighted in the comparison of $M_\mathrm{B+D}$ and $M_\mathrm{B}+M_\mathrm{D}$ discussed in Section \ref{mass_comp}.  In Figure \ref{fig:bd_nodust} we show the comparison between $M_\mathrm{B+D}$ and $M_\mathrm{B} + M_\mathrm{D}$ as a function of $(B/T)_r$, analogous to the right-hand panel of Figure \ref{fig:mass_comp}, for the dust-free masses.  Contrary to the comparison in Figure \ref{fig:mass_comp}, the dust-free fits remain relatively well behaved even at low $B/T$, and likely provide a more robust estimate of the bulge mass in these systems.  The behavior of disks in the dusty vs. dust-free fits is similar to that of the total photometry: the reddest disks get slightly lighter in the dust-free fits, while star-forming disks get slightly heavier.

\subsection{Comparison to the MPA-JHU stellar masses}

\begin{figure*}
\centering
\includegraphics[scale=0.95]{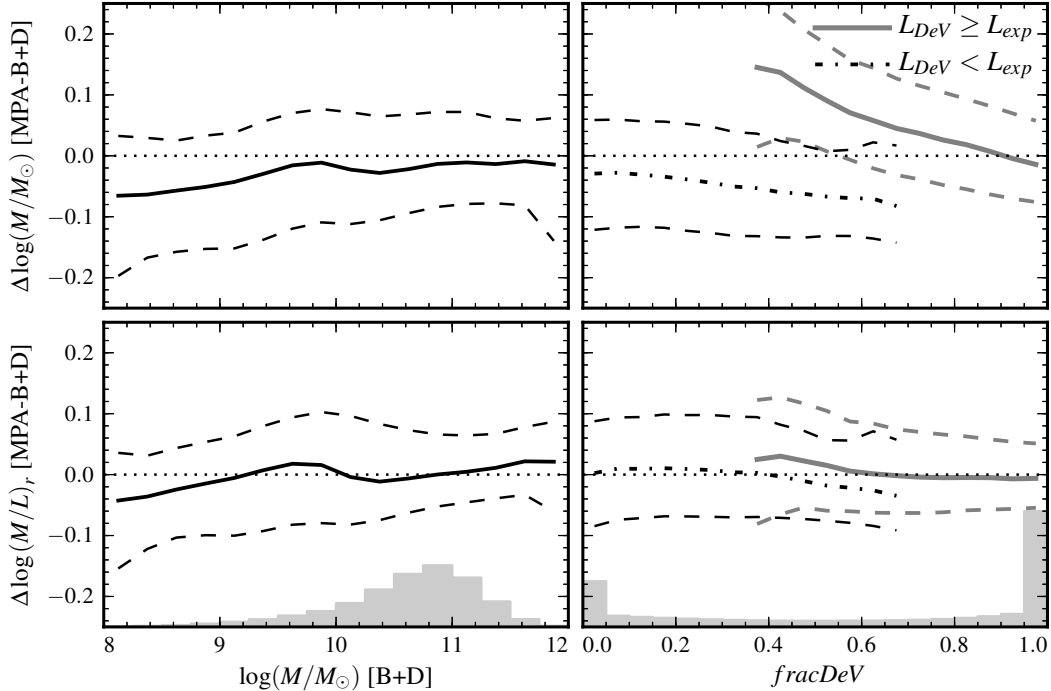}
\caption{Comparison of stellar mass and \emph{M/L} between our fits and the masses available in the MPA-JHU data catalogs, where $\Delta$ is determined as the MPA-JHU value minus our value.  In the left panels, solid lines show the median offset in either total stellar mass (top panel) or $r$-band \emph{M/L} as a function of stellar mass; dashed lines show the 16th and 84th percentile values of the distribution.  In the right panels we show a similar comparison between total mass and \emph{M/L}, this time as a function of \emph{fracDeV}.  Here we have split the galaxy population into galaxies that are best fit by a de Vaucouleurs profile (solid red lines) or an exponential profile (dot-dashed blue lines) according to the SDSS pipeline.  The shaded histograms in the bottom panels show the relative distribution of galaxies in either stellar mass (left) or \emph{fracDeV} (right).}
\label{fig:mpa_comp}
\end{figure*}

The MPA-JHU group provide a catalog of stellar masses which have been widely used in analyses of the SDSS\footnote{available at \url{http://www.mpa-garching.mpg.de/SDSS/DR7/}} \citep{kauffmann2003b,brinchmann2004,salim2007}.  These data therefore provide a well-tested reference point for the mass estimates presented here.  It is important to note that there are two effects at play in these comparisons: the first reflects differences between the SDSS- and {\sc gim2d}-derived photometry, which are discussed extensively in \citetalias{simard2011}, while the second relates to differences expected as a result of variations in the underlying SPS models.  Masses derived in the MPA-JHU catalogs for DR7 are based on SED fits to the SDSS {\sc model} magnitudes, which are estimated from parametric fits to the observed surface-brightness profile.  In each case, galaxies are fitted separately with exponential and de Vaucouleurs profiles, and the relative likelihood of these fits is used to determine which profile provides the most appropriate description of the data.  The final {\sc model} magnitudes are derived by integrating the best-fit (de Vaucouleurs or exponential) profile out to large radii.  The underlying SPS models are computed using the {\sc galaxev} code \citep{bruzual2003} and a Monte Carlo approach to construct a large suite of templates with varying star-formation histories, formation times, and metallicities.  As discussed in Section \ref{models}, one of the most significant structural differences between the MPA-JHU SPS library and our own is their inclusion of bursts at late times, which results in relatively small differences in the derived stellar mass for most galaxies.  It is beyond the scope of the present work to address the effects of this difference in detail, and we refer the reader to \citet{gallazzi2009} and \citet{walcher2011} for a more extensive discussion of this issue.

In the left panels of Figure \ref{fig:mpa_comp} we plot the median difference in total stellar mass (top panel) and \emph{M/L} (bottom panel) between the MPA-JHU mass estimates and our own as a function of stellar mass from the bulge+disk profile fits.  We find that masses computed using {\sc gim2d} photometry and FSPS are $\sim$0.02 dex heavier on average than the MPA-JHU mass estimates, with slightly larger deviations for low-mass galaxies.  Comparing the top and bottom panels, we see that the offset in total mass generally exceeds that of \emph{M/L}, which is a consequence of the higher average flux derived by \citetalias{simard2011} relative to the SDSS {\sc model} magnitudes.  At low masses, offsets in total mass are almost entirely accounted for by variation in \emph{M/L}, suggesting that the underlying models may play a more important role in star-forming galaxies.  As discussed in Section \ref{models}, application of smoothly-evolving models---such as those used here---to bursty systems can lead to a systematic overestimate of their \emph{M/L}. The increased \emph{M/L} relative to the MPA-JHU catalog may therefore reflect a combination of differences in the two models grids and the increasingly stochastic nature of star-formation in low-mass galaxies; however, we see no evidence that this results in a significant offset in the derived \emph{M/L} on average. The comparison discussed above is qualitatively similar if we use masses determined from the S\'ersic photometry, which are on average $\sim$0.08 dex heavier than the MPA-JHU masses.  At high mass the S\'ersic-profile based masses deviate more significantly from the MPA-JHU masses, and are up to 0.2 dex heavier at $\log (M/M_\odot) = 12$.  This same behavior is shown in Figure \ref{fig:mass_comp} in comparison to the bulge+disk masses.  

The discussion above is focused primarily on galaxy properties (mass or \emph{M/L}) as a function of their stellar mass; however, we have largely neglected details of the photometric measurements that can significantly affect the recovery of galaxies' stellar mass.  For example, both \citetalias{simard2011} and \citet{taylor2011} have shown that the SDSS {\sc model} magnitudes have a tendency to bifurcate for galaxies with intermediate structural properties (i.e. $0.25 < B/T < 0.6$ or $1.5 < n < 3$), resulting in a spurious bimodality in the derived photometry for such systems.  This structural dependence can manifest itself as a systematic shift in the derived stellar masses of objects that host both bulges and disks \citep[e.g][]{taylor2011}.  In the right panels of Figure \ref{fig:mpa_comp} we investigate this structural dependence explicitly in our data by comparing stellar mass and \emph{M/L} as a function of \emph{fracDeV}, the fraction of light contained within the de Vaucouleurs component fit by the SDSS {\sc photo} pipeline.  In both the top and bottom panels, the median offsets for galaxies with {\sc model} magnitudes derived from the de Vaucouleurs (exponential) fits are shown by the solid (dot-dashed) lines.  Here we can see that galaxies with intermediate \emph{fracDeV} can have their masses over- or underestimated by $\sim$0.2 dex, but that there is good agreement between \emph{M/L} estimates for any given \emph{fracDeV}.  The disagreement between the masses presented here and those in the MPA-JHU catalog at intermediate \emph{fracDeV} is primarily due to fitting a single, fixed profile (de Vaucouleurs or exponential) as opposed to something that can better accommodate complex structure; any differences between the SED fitting procedures appear to be small once an appropriate model is fit to the galaxy light profile.  Looking at it another way, there is relatively good agreement between the {\sc model} and {\sc gim2d} colors, but the total fluxes can vary significantly depending on galaxies' underlying structure.

\section{Systematic uncertainties}
\label{uncertainties}

In Sections \ref{data}, \ref{mass_estimates}, and Appendix \ref{flux_sims} we have already discussed some of the potential uncertainties present in the photometric catalogs on which our mass estimates are based.  We now attempt a more direct discussion of uncertainties in the SED fitting procedure itself.  Our goal in this Section is to provide would-be users with a sense of the uncertainties present in our stellar mass estimates and, in fact, many mass estimates based on broadband SEDs.  We have chosen to order this discussion in a top down fashion, starting with the most significant sources of uncertainty and working down towards the finer details.

\begin{figure}
\centering
\includegraphics[scale=1.0]{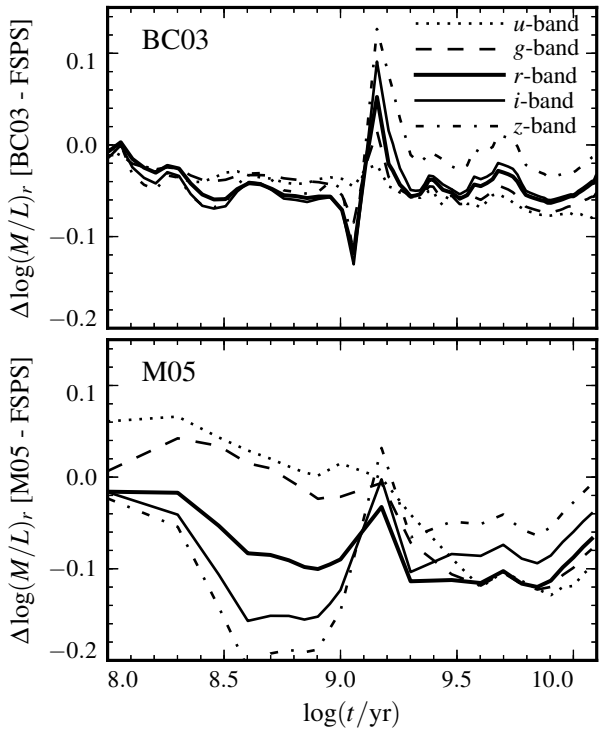}
\caption{Predicted \emph{M/L} relative to FSPS for the \citetalias{bruzual2003} and \citetalias{maraston2005} SPS models described in Section \ref{model_comp}.  In all cases we plot the change in \emph{M/L} as either \citetalias{bruzual2003} or \citetalias{maraston2005} minus the FSPS value.}
\label{fig:model_comp}
\end{figure}

\subsection{The choice of SPS code}
\label{model_comp}

There are usually multiple ways to approach any problem, and stellar population synthesis is no exception.  While we have chosen to use FSPS as the fiducial model in our mass estimates, there are many other SPS models based on a variety of stellar evolutionary tracks and spectral libraries that could have been used to the same end \citep[e.g.][]{fioc1997,leitherer1999,bruzual2003,maraston2005,vazdekis2012}.  It is therefore crucial to understand the level at which these different models produce consistent predictions, and the extent to which (and where) they disagree.

In Figure \ref{fig:model_comp} we show the variation in $ugriz$ \emph{M/L}s relative to FSPS as a function of population age for two different SPS models: \citet[top panel, hereafter \citetalias{bruzual2003}]{bruzual2003} and \citet[bottom panel, hereafter \citetalias{maraston2005}]{maraston2005}.  In all cases we show the results for a dust-free single stellar population with solar metallicity.    \emph{M/L} values from the \citetalias{bruzual2003} models are computed using a \citet{chabrier2003} IMF, and so are directly comparable to the FSPS models.  \citetalias{maraston2005} models are based on a \citet{kroupa2001} IMF, which we convert to \citet{chabrier2003} using an offset of 0.05 dex \citep[Section \ref{IMF_extinction};][]{bernardi2010}.  The \citetalias{bruzual2003} models are based on an earlier version of the Padova isochrones than FSPS \citep{girardi1996}, and the luminosity at all phases of stellar evolution is computed via an isochrone synthesis technique.  \citetalias{maraston2005} models are based on isochrones and evolutionary tracks from \citet{cassisi1997}, and use a similar approach to \citetalias{bruzual2003} and FSPS to compute the luminosity of stars up to the main sequence turnoff.  For evolved stars (post-main sequence), \citetalias{maraston2005} compute luminosities via fuel consumption theory rather than isochrone synthesis.  The most significant difference between the \citetalias{maraston2005} and \citetalias{bruzual2003} or FSPS models is that the luminosity of thermal-pulsing AGB (tp-AGB) stars derived from fuel consumption theory is higher relative to other calculations.

On average the \emph{M/L}s predicted by FSPS are higher by 0.04--0.1 dex (10-25\%)---that is, the models have a higher stellar mass per unit luminosity---relative to either the \citetalias{bruzual2003} or \citetalias{maraston2005} models.  This systematic offset notwithstanding, the FSPS and \citetalias{bruzual2003} models agree reasonably well at all wavelengths with a typical spread of 0.04-0.06 dex that increases slightly towards old ages.  

The comparison with \citetalias{maraston2005} models is less favorable; however, if we restrict the comparison to relatively old ages ($\gtrsim 2$ Gyrs) then the scatter is generally less than 0.1 dex.  At ages less than 1 Gyr there is significant spread in the relative \emph{M/L} depending on wavelength: there is relatively good agreement between the $u$- and $g$-band \emph{M/L}s, while the $i$- and $z$-band \emph{M/L}s are up to 0.2 dex lower in \citetalias{maraston2005} relative to FSPS.  This decrease in \emph{M/L} can be traced to the increased luminosity of tp-AGB stars in the \citetalias{maraston2005} models relative either FSPS or \citetalias{bruzual2003}, which contribute significantly in the near-IR.  We will discuss the influence of tp-AGB stars more thoroughly in Section \ref{stellar_evo}, and here simply note that they represent a significant uncertainty in the determination of stellar masses.  In both comparisons, the feature at $\sim$1 Gyr is due to differences in the onset of core helium burning in the stellar evolutionary tracks used by the different models.

The above comparison points to an inherent (and unavoidable!) uncertainty in \emph{any} stellar mass estimates based on broadband photometry of 0.1-0.2 dex, depending on the properties of the underlying stellar population.  This is in relatively good agreement with the results of previous studies which have addressed the differences between various SPS models \citep[e.g.][]{bell2001,maraston2005,marchesini2009,ilbert2010,swindle2011}.

\subsection{The IMF and extinction law}
\label{IMF_extinction}

Among the assumptions that go into the construction of our SPS grid, changes in the stellar IMF and extinction law can lead to significant systematic shifts in galaxies' derived stellar mass, and are therefore worth a moment of discussion.

In addition to the relatively large uncertainty on the slope of the IMF locally \citep{kroupa2001}, over the past several years there has been mounting evidence for systematic variation of the stellar IMF as a function of galaxy properties \citep[e.g.][]{cenarro2003,hoversten2008,van-dokkum2010a,gunawardhana2011,dutton2012}.  While the exact form of this variation is still poorly understood, it seems that some variability in either or both the high and low mass IMF slope should be considered when discussing the overall influence of IMF uncertainties on our final mass catalog.  

Changes in the IMF slope at $M_\odot \lesssim 1$ can be approximated by a simple offset in \emph{M/L} (or stellar mass) for all galaxies, as low-mass stars contribute relatively little to galaxies' overall luminosity.  Of the IMFs most frequently adopted in the literature, those of \citet{salpeter1955}, \citet{kroupa2001}, and \citet{chabrier2003} have similar logarithmic slopes at high mass, and can therefore be transformed in this fashion.  Roughly speaking, masses derived using a \citet{salpeter1955} or \citet{kroupa2001} IMF are 0.25 and 0.05 dex \emph{heavier} than those estimated assuming a \citet{chabrier2003} IMF.  On the other hand, changing the slope of the IMF at $M_\odot \gtrsim 1$ has a significant impact on the rate at which the luminosity of a stellar population evolves \citep{tinsley1980}.  Therefore, unlike changing the IMF for low-mass stars, changing the slope of the IMF at high mass results in changes in \emph{M/L} which depend on the (luminosity-weighted) age of the stellar population.  To place this in the context of our mass estimates, adopting a power-law IMF slope of $x = 1.0$ for $M_\odot > 1$ (in units where a \citealp{chabrier2003} IMF has $x = 1.3$) leads to an increase in the \emph{M/L} of red galaxies of $\sim$0.1 dex, but relatively little change in the \emph{M/L} of blue galaxies.  Similarly, adopting $x = 1.6$ leads to no change in \emph{M/L} for red galaxies, but a 0.05 dex increase for blue galaxies.

We showed in Section \ref{dust_free} that removing \ebv as a free parameter can lead to significant changes in the derived mass, particularly when models are poorly constrained by the photometric data. Even when considering dust properties as a variable, the particular choice of extinction law can have a significant effect on the derived masses.  In constructing our fiducial grid we have used the \citet{calzetti2000} extinction curve, however empirical attenuation laws based on the Milky Way \citep[MW; e.g.,][]{cardelli1989} and Small Magellanic Cloud \citep[SMC; e.g.,][]{pei1992}, as well as single- or multi-component power-law curves \citep[e.g.][]{charlot2000} are also common in the literature.  Changing among the \citet{calzetti2000}, MW, and SMC extinction curves leads to variations in total mass of 0.1 to 0.2 dex depending on the wavelengths used in the SED fitting procedure \citep[e.g.][]{marchesini2009,ilbert2010,swindle2011,pforr2012}; differences are typically largest when incorporating data which extend into the far-UV.  The dust model of \citet{charlot2000} includes additional attenuation for young objects to allow for the effects of stars embedded in HII regions, and can lead to even more significant variation for particular subsamples of (primarily young) galaxies \citep[e.g.][]{ilbert2010,swindle2011}.  At optical wavelengths, however, variations in the extinction law lead to uncertainties of $\sim$0.15 dex.

\subsection{Uncertainties in stellar evolution}
\label{stellar_evo}

We now turn to a consideration of systematic effects as a result of uncertain phases of stellar evolution, focusing primarily on the contributions of blue horizontal branch (BHB), blue straggler (BS), and thermally-pulsing asymptotic giant branch (tp-AGB) stars.

In the context of FSPS, variations in horizontal branch (HB) morphology are parameterized by the fraction of HB stars bluer than the red clump.  These stars are assumed to uniformly populate the horizontal branch between the temperature of the red clump and 10,000 K.  We adopt two values for the fraction of blue HB stars, $f_\mathrm{BHB}$, of 0 and 20\%, which encompass the abundance of very blue stars necessary to explain the UV upturn in elliptical galaxies \citep{dorman1995}.  We note that our fiducial model includes a 5\% contribution by mass in metal poor stars ($\left[\mathrm{Z/H}\right] \approx -2$), and therefore a minimum 5\% contribution of BHB stars might be more appropriate; however, in order to facilitate comparison with previous SPS models which do not consider contributions from BHB stars \citep[e.g.][]{bruzual2003} we have elected to adopt a minimum model with $f_\mathrm{BHB} = 0$.  BS stars are defined in terms of the specific frequency relative to HB stars, $S_\mathrm{BS}$, and populate a parameter space between 0.5 and 2.5 mag brighter than the zero-age main sequence.  We consider models in which $S_\mathrm{BS}$ is either 0 or 2.  Both BHB and BS stars are assumed to contribute only in populations older than 5 Gyrs.

Luminous tp-AGB stars contribute significantly to flux in the near-infrared, particularly in populations with ages of $\sim$1--2 Gyrs, and are therefore an important consideration for the estimation of stellar mass in such galaxies.  Currently, the overall contribution of tp-AGB stars is extremely uncertain.  On the one hand, the stellar populations of intermediate-age star clusters seem to require a significant contribution from tp-AGB stars to describe their observed photometric and spectral properties \citep[e.g.][]{maraston2005,lyubenova2010,lyubenova2012}.  However, the extent to which such stars contribute to the SEDs of intermediate-age galaxies is less clear \citep[e.g.][]{kriek2010,conroy2010b,zibetti2013}.  FSPS allows for shifts in the characteristic temperature $T$ and luminosity $L$ of tp-AGB stars, and we adopt a relatively broad range of $\Delta\log L = \pm0.4$ dex and $\Delta\log T = \pm0.2$ dex, reflecting the significant uncertainty of this evolutionary phase.  The fiducial FSPS models already incorporate an empirical adjustment to the tp-AGB temperature and luminosity, described by \citet{conroy2010b}.  The values of \dell and \delt used here characterize shifts in tp-AGB luminosity and temperature relative to the already-calibrated FSPS models.

\begin{figure}
\centering
\includegraphics[scale=0.80]{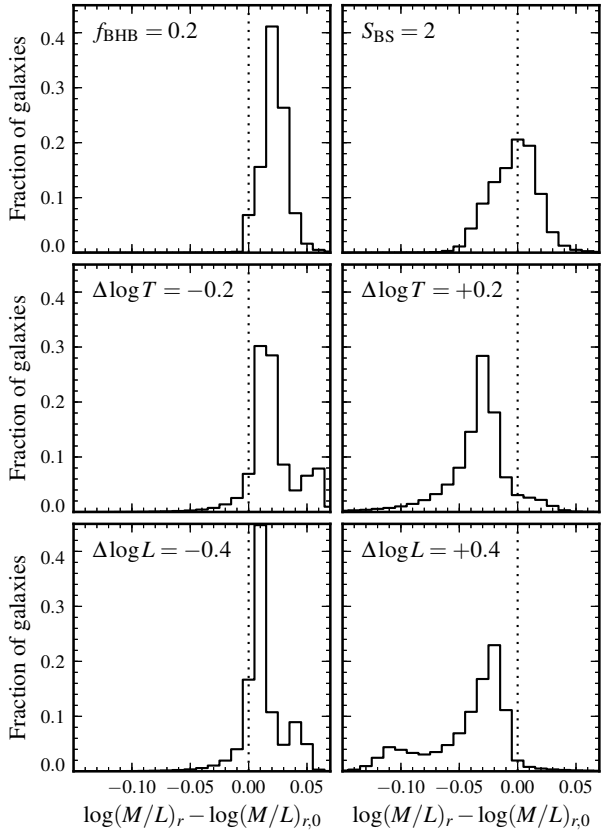}
\caption{Offset between our fiducial SPS model and models that include variations in the accounting of BHB, BS, and tp-AGB stars.  Vertical dotted lines represent no offset.}
\label{fig:delta_ml}
\end{figure}

In Figure \ref{fig:delta_ml} we plot the offset in $r$-band \emph{M/L} relative to our fiducial model for various contributions of BHB, BS and tp-AGB stars as indicated in each panel.  Dotted line represent no shift in \emph{M/L}, and are shown for reference only.  Figure \ref{fig:delta_ml} suggests that varying stellar evolutionary phases result in relatively small systematic deviations in the derived \emph{M/L}, typically less than $\sim$0.05 dex, in good agreement with the findings of \citet{conroy2009}.  Both BHB and BS stars appear to play only a minor part in the overall systematic uncertainty on the stellar masses, though the inclusion of 20\% BHB stars can result in a shift of the derived \emph{M/L} of 0.02--0.03 dex.  Uncertainties in the tp-AGB phase introduce a much larger spread in the mass estimates, particularly given increases in the tp-AGB luminosity (bottom right panel of Figure \ref{fig:delta_ml}).  This is consistent with the SPS model comparison discussed in Section \ref{model_comp}.  It is important to note, however, that the relative improvement in the \emph{quality} of the model fits (as judged by their $\chi^2$) is minimal for the adopted variations in both \dell and \delt.  As discussed by \citet{conroy2009}, the primary effect of treating \dell and \delt as free parameters is to increase the uncertainty in \emph{M/L}, but not necessarily change its value systematically.  In this sense, the offsets shown in Figure \ref{fig:delta_ml} represent upper limits on the expected variation of mass given the adopted parameter range.

\subsubsection{Color dependence}

In addition to the globally-averaged uncertainties in \emph{M/L} discussed above, it is equally important to understand how these uncertainties vary as a function of galaxy properties.

\begin{figure*}
\centering
\includegraphics[scale=0.90]{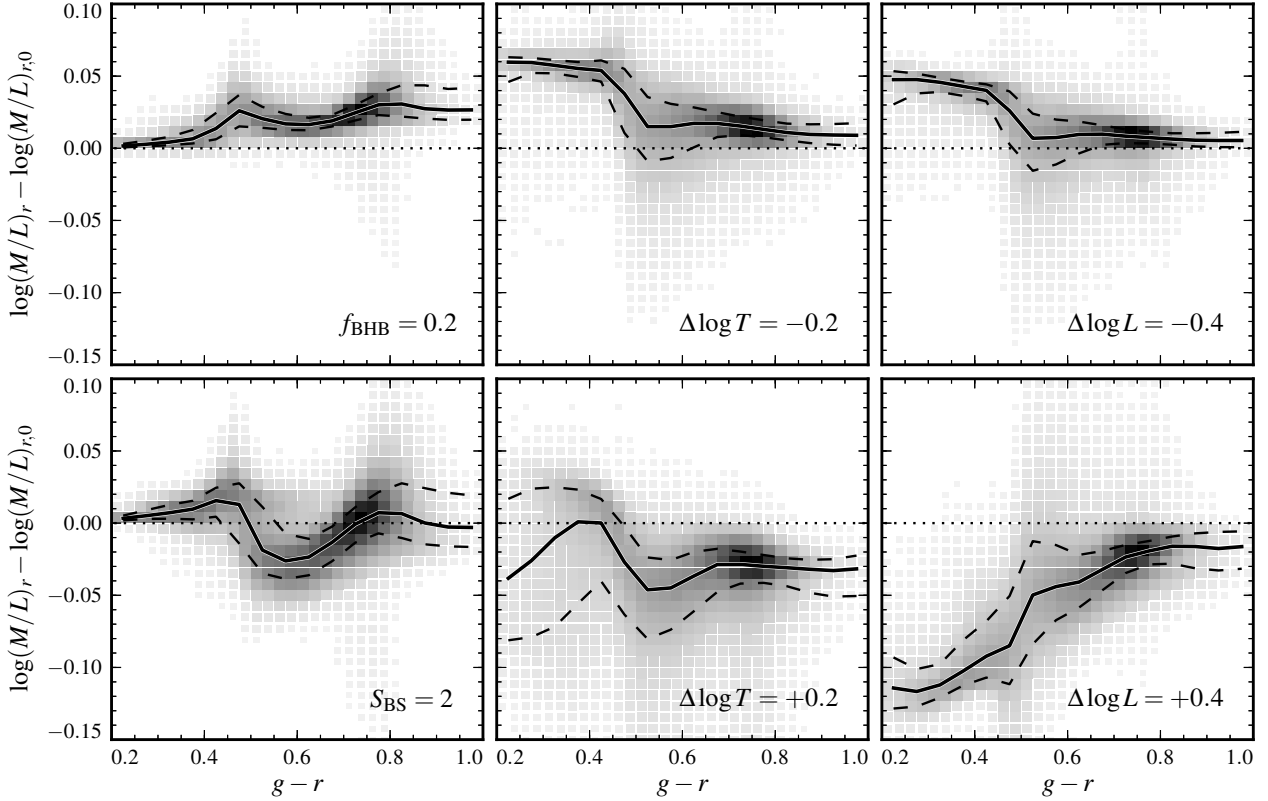}
\caption{$r$-band \emph{M/L} offset as a function of galaxy color for different stellar population models.  In each panel, solid lines indicate the median \emph{M/L} offset as a function of color, while dashed lines show the $\pm$1$\sigma$ confidence interval.  Shading shows the underlying galaxy distribution.  Dotted lines mark no offset and are shown for reference only.}
\label{fig:ml_color}
\end{figure*}

In Figure \ref{fig:ml_color} we show the change in stellar \emph{M/L} as a function of galaxies' \gr color for the same models considered in Figure \ref{fig:delta_ml}.  In each panel, solid lines show the median offset in \emph{M/L}, while dashed lines show the $\pm$1 sigma percentile confidence intervals.  Shading shows the underlying galaxy distribution.  

The systematic offsets shown in Figure \ref{fig:delta_ml} are echoed in Figure \ref{fig:ml_color}, but it highlights the dependence of these systematic effects on the underlying stellar population.  The effects of both the BHB and BS contributions are relatively minor, at most $\sim$0.05 dex.  While there is no obvious systematic trend for BS stars, including 20\% BHB stars can lead to an increase in the recovered stellar mass by 0.02--0.03 dex for red galaxies relative to blue.  Recall that BHB (and BS) stars are only implemented in populations older than 5 Gyr, so that this color dependence reflects the correlation between observed color and mean (luminosity-weighted) age of the stellar population.  In part, the lack of sensitivity to BS and BHB stars in our fits can be attributed to the large observational uncertainties on $u$-band flux estimates relative to other bandpasses.

As discussed in previous sections, uncertainties in the properties of tp-AGB stars are significant, and these uncertainties manifest themselves most clearly in young stellar populations.  The relative sensitivity of a given galaxy's recovered stellar mass to the temperature and luminosity of tp-AGB stars therefore hinges on the average age of its stellar population, and is emphasized by the strong dependence of recovered mass on galaxy color in Figure \ref{fig:ml_color}.  For reference, the division between red sequence and blue cloud galaxies falls at $g-r \sim 0.4-0.5$.  For blue galaxies, variations in \dell and \delt can lead to systematic shifts in $r$-band \emph{M/L} of up to 0.1 dex for most galaxies, and up to 0.2--0.3 dex in extreme cases.  Conversely, older (redder) galaxies are relatively insensitive to the properties of tp-AGB stars, as RGB stars dominate the flux output of post main sequence stars in populations more than 2--3 Gyrs old.  Recent observational evidence suggests that the lifetimes of low-mass AGB stars may be significantly overestimated \citep{girardi2010,melbourne2012}, and therefore the total flux from tp-AGB stars may be too high by a factors of a few in the models of \citet{marigo2007}.  These findings are supported by the comparison of post-starburst galaxy spectra with synthetic infrared SPS SEDs, which point to a relatively small contribution from tp-AGB stars even in young galaxy populations \citep[e.g.][]{kriek2010,zibetti2013}.  As mentioned previously, FSPS already incorporates an empirical calibration of \dell and \delt relative to the fiducial tp-AGB evolutionary tracks of \citet{marigo2007}, described by Equations 2 and 3 of \citet{conroy2010b}.  These corrections range from \dell = $-1.0$ to $-0.2$ and \delt = $+0.1$ to $-0.1$ depending on the age and metallicity of the stellar population.  \citet{girardi2010} show that the evolutionary tracks of \citet{marigo2007} may overestimate the frequency of AGB stars by factors of 3 to 7, naively suggesting that at \dell of $-0.3$ to $-0.8$ may be necessary to match the observed data, within the range of the FSPS calibration.  \citet{kriek2010} find that a similar offset  ($\Delta \log L \approx -0.6$) is required to fit their sample of post-starburst galaxies in the NEWFIRM medium-band survey.  Therefore, while adopting $\Delta \log L = 0.4$ leads to a significant differential offset between young and old galaxies of 0.1 dex, empirical evidence suggests that typical tp-AGB luminosities are unlikely to be underestimated (and may, in fact, be overestimated).

\section{How to use these catalogs}
\label{discussion}

In the preceding Sections we have described the mechanics of our stellar mass computations, and have attempted to highlight the uncertainties that should be borne in mind with their use.  We now turn to a discussion of more practical matters.

Estimates of stellar mass computed using the full SPS grid are given in Tables \ref{sersic_mass_table} and \ref{bd_mass_table} for single (S\'ersic) and two-component (bulge+disk) profile fits, respectively.  Masses computed using dust-free models are given in Tables \ref{sersic_mass_table_nodust} and \ref{bd_mass_table_nodust}.  Along with the mass estimates we include several other parameters---\zmin, \zmax, profile type, $F$-test probability, and bulge+disk mass offset---described in the following Sections.  In instances where there is no flux in all five of the $ugriz$ bands, for example where the system is a "pure" bulge or disk, masses have been given a value of -99.99.

\begin{deluxetable*}{lcrrrrrr}
\tablewidth{0pc}
\tablecaption{Total stellar mass estimates from S\'ersic photometry}
\tablehead{\colhead{objID} & \colhead{$z$} & \colhead{$\log M$} & \colhead{$\log M_{\mathrm{p16}}$} & \colhead{$\log M_{\mathrm{p84}}$} & \colhead{$z_\mathrm{min}$} & \colhead{$z_\mathrm{max}$} & \colhead{Type}\\
\colhead{(1)} & \colhead{(2)} & \colhead{(3)} & \colhead{(4)} & \colhead{(5)} & \colhead{(6)} & \colhead{(7)} & \colhead{(8)}
}
\startdata
587724242842026102 & 0.0139 & 10.32 & 10.20 & 10.41 & 0.0137 & 0.0729 & 3\\
587724242303713336 & 0.0134 & 10.03 & 9.88 & 10.13 & 0.0121 & 0.0653 & 3\\
587728669882515508 & 0.0150 & 10.09 & 9.94 & 10.16 & 0.0122 & 0.0657 & 3\\
587729386602692631 & 0.0105 & 10.00 & 9.83 & 10.09 & 0.0096 & 0.0514 & 3\\
587722981745295552 & 0.0234 & 10.05 & 9.95 & 10.30 & 0.0140 & 0.0753 & 3\\
587722952767439282 & 0.0337 & 10.13 & 9.93 & 10.27 & 0.0130 & 0.0694 & 2\\
587722981747392587 & 0.0187 & 10.09 & 9.85 & 10.20 & 0.0153 & 0.0825 & 2\\
587722953841442929 & 0.0405 & 10.58 & 10.44 & 10.67 & 0.0204 & 0.1035 & 3\\
587722953841770855 & 0.0386 & 10.59 & 10.35 & 10.69 & 0.0232 & 0.1182 & 3\\
587722953303916807 & 0.0333 & 10.56 & 10.41 & 10.65 & 0.0213 & 0.1077 & 4\\
587722981742936247 & 0.0420 & 10.02 & 9.77 & 10.21 & 0.0160 & 0.0865 & 2\\
587722953304113519 & 0.0333 & 10.03 & 9.84 & 10.15 & 0.0138 & 0.0749 & 3\\
587722981748048047 & 0.0405 & 10.05 & 9.81 & 10.16 & 0.0129 & 0.0696 & 2\\
587722953304310243 & 0.0346 & 10.17 & 10.02 & 10.26 & 0.0122 & 0.0654 & 3\\
587722981749227687 & 0.0445 & 10.55 & 10.40 & 10.78 & 0.0247 & 0.1237 & 3\\
587722982283214886 & 0.0231 & 10.25 & 9.99 & 10.39 & 0.0134 & 0.0714 & 2\\
587722981750014081 & 0.0373 & 10.36 & 10.12 & 10.47 & 0.0185 & 0.0969 & 3\\
587722982284460272 & 0.0191 & 10.48 & 10.34 & 10.59 & 0.0175 & 0.0903 & 3\\
587722982278365259 & 0.0350 & 10.32 & 10.20 & 10.47 & 0.0185 & 0.0962 & 3\\
587722953304572282 & 0.0336 & 10.15 & 9.94 & 10.25 & 0.0138 & 0.0746 & 3\\
\enddata
\tablecomments{Columns: (1) SDSS object ID, (2) spectroscopic redshift, (3) total stellar mass, (4) 16th percentile of the total mass PDF, (5) 84th percentile of the total mass PDF, (6) minimum redshift at which the galaxy satisfies our sample selection criteria, (7) maximum redshift at which the galaxy satisfies our sample selection criteria, (8) profile type. (This table is available in its entirety in a machine-readable form in the online journal.  A portion is shown here for guidance regarding its form and content.)\label{sersic_mass_table}} 
\end{deluxetable*}

\begin{turnpage}
\begin{deluxetable*}{lcrrrrrrrrrrrrrr}
\tablewidth{0pc}
\tablecaption{Total, bulge, and disk stellar mass estimates from bulge+disk photometry}
\tablehead{\colhead{objID} & \colhead{$z$} & \colhead{$\log M$} & \colhead{$\log M_{\mathrm{p16}}$} & \colhead{$\log M_{\mathrm{p84}}$} & \colhead{$\log M_\mathrm{B}$} & \colhead{$\log M_{\mathrm{B},\,\mathrm{p16}}$} & \colhead{$\log M_{\mathrm{B},\,\mathrm{p84}}$} & \colhead{$\log M_\mathrm{D}$} & \colhead{$\log M_{\mathrm{D},\,\mathrm{p16}}$} & \colhead{$\log M_{\mathrm{D},\,\mathrm{p84}}$} & \colhead{$z_\mathrm{min}$} & \colhead{$z_\mathrm{max}$} & \colhead{\pps} & \colhead{Type} & \colhead{$\Delta_\mathrm{B+D}$}\\
\colhead{(1)} & \colhead{(2)} & \colhead{(3)} & \colhead{(4)} & \colhead{(5)} & \colhead{(6)} & \colhead{(7)} & \colhead{(8)} & \colhead{(9)} & \colhead{(10)} & \colhead{(11)} & \colhead{(12)} & \colhead{(13)} & \colhead{(14)} & \colhead{(15)} & \colhead{(16)}
}
\startdata
587735348040433783 & 0.0067 & 11.17 & 11.14 & 11.18 & 9.33 & 8.89 & 9.54 & 11.14 & 11.11 & 11.15 & 0.0050 & 0.0245 & 0.487 & 3 & 0.65\\
587722981745295552 & 0.0234 & 10.12 & 9.93 & 10.37 & 10.15 & 10.03 & 10.23 & 9.83 & 9.72 & 10.05 & 0.0140 & 0.0754 & 0.004 & 3 & 0.94\\
587722952230371590 & 0.0618 & 10.07 & 9.86 & 10.22 & 9.52 & 9.30 & 9.67 & 9.79 & 9.66 & 9.91 & 0.0130 & 0.0699 & 0.500 & 4 & 0.43\\
587722952230175145 & 0.0716 & 10.51 & 10.32 & 10.65 & 9.57 & 9.33 & 9.79 & 10.47 & 10.28 & 10.60 & 0.0225 & 0.1154 & 0.040 & 2 & 0.05\\
587722952230175173 & 0.0720 & 10.59 & 10.39 & 10.76 & 11.16 & 11.06 & 11.23 & 10.03 & 9.95 & 10.17 & 0.0249 & 0.1258 & 0.183 & 3 & 3.59\\
587722953841312162 & 0.0628 & 10.75 & 10.59 & 10.87 & 10.50 & 10.33 & 10.61 & 10.45 & 10.30 & 10.56 & 0.0260 & 0.1293 & 0.043 & 3 & 0.16\\
587722981737431161 & 0.0743 & 10.38 & 10.18 & 10.52 & -99.99 & -99.99 & -99.99 & 10.40 & 10.20 & 10.54 & 0.0179 & 0.0938 & 0.748 & 2 & 0.09\\
587722981741625520 & 0.0706 & 10.69 & 10.54 & 10.76 & 10.52 & 10.35 & 10.63 & 10.25 & 10.11 & 10.36 & 0.0256 & 0.1269 & 0.459 & 3 & 0.14\\
587722981747392587 & 0.0187 & 10.05 & 9.79 & 10.18 & 8.92 & 8.69 & 9.09 & 10.06 & 9.82 & 10.21 & 0.0153 & 0.0827 & 1.000 & 2 & 0.16\\
587722981746016481 & 0.0575 & 10.00 & 9.89 & 10.20 & 9.86 & 9.67 & 10.02 & 9.88 & 9.77 & 10.02 & 0.0190 & 0.1011 & 0.392 & 3 & 0.91\\
587722952230240617 & 0.0548 & 10.46 & 10.28 & 10.71 & 11.03 & 10.94 & 11.11 & 10.11 & 10.04 & 10.23 & 0.0312 & 0.1579 & 0.244 & 3 & 3.65\\
587722981742149718 & 0.0700 & 10.73 & 10.52 & 10.82 & 10.45 & 10.31 & 10.57 & 10.48 & 10.33 & 10.60 & 0.0316 & 0.1591 & 0.187 & 3 & 0.23\\
587722952767439282 & 0.0337 & 10.17 & 9.99 & 10.30 & 8.64 & 8.30 & 8.89 & 10.18 & 10.00 & 10.29 & 0.0130 & 0.0694 & 0.779 & 2 & 0.07\\
587735349648752750 & 0.0230 & 10.23 & 10.00 & 10.39 & 10.13 & 10.02 & 10.22 & 9.90 & 9.80 & 10.10 & 0.0168 & 0.0896 & 0.015 & 3 & 0.51\\
587722953841770855 & 0.0386 & 10.58 & 10.34 & 10.68 & 10.40 & 10.29 & 10.50 & 10.27 & 10.13 & 10.44 & 0.0232 & 0.1183 & 0.062 & 3 & 0.35\\
587722981747916880 & 0.0637 & 10.77 & 10.60 & 10.94 & 10.74 & 10.59 & 10.82 & 10.33 & 10.18 & 10.48 & 0.0308 & 0.1498 & 0.048 & 3 & 0.61\\
587722981742215240 & 0.0710 & 10.33 & 10.16 & 10.44 & 10.10 & 9.97 & 10.21 & 10.02 & 9.82 & 10.17 & 0.0169 & 0.0889 & 0.464 & 3 & 0.19\\
587722952230961709 & 0.0543 & 10.14 & 9.99 & 10.24 & 10.16 & 10.00 & 10.27 & 8.82 & 8.48 & 9.18 & 0.0134 & 0.0720 & 0.516 & 4 & 0.21\\
587722981742936247 & 0.0420 & 10.13 & 9.87 & 10.27 & -99.99 & -99.99 & -99.99 & 10.16 & 9.94 & 10.28 & 0.0160 & 0.0865 & 0.999 & 2 & 0.13\\
587722953303916807 & 0.0333 & 10.53 & 10.39 & 10.59 & 10.45 & 10.29 & 10.52 & 9.94 & 9.78 & 10.07 & 0.0213 & 0.1080 & 0.353 & 4 & 0.24\\
\enddata
\tablecomments{Columns: (1) SDSS object ID, (2) spectroscopic redshift, (3) total stellar mass, (4) 16th percentile of the total mass PDF, (5) 84th percentile of the total mass PDF, (6) bulge stellar mass, (7) 16th percentile of the bulge stellar mass PDF, (8) 84th percentile of the bulge mass PDF, (9) disk stellar mass, (10) 16th percentile of the disk stellar mass PDF, (11) 84th percentile of the disk mass PDF, (12) minimum redshift at which the galaxy satisfies our sample selection criteria, (13) maximum redshift at which the galaxy satisfies our sample selection criteria, (14) $F$-test probability that two components are \emph{not} required, (15) profile type, (16) difference between the total mass and sum of the bulge and disk masses in units of the standard error. (This table is available in its entirety in a machine-readable form in the online journal.  A portion is shown here for guidance regarding its form and content.)\label{bd_mass_table}} 
\end{deluxetable*}	
\end{turnpage}

\begin{deluxetable*}{lcrrrrrr}
\tablewidth{0pc}
\tablecaption{Total stellar mass estimates from S\'ersic photometry with dust-free models}
\tablehead{\colhead{objID} & \colhead{$z$} & \colhead{$\log M$} & \colhead{$\log M_{\mathrm{p16}}$} & \colhead{$\log M_{\mathrm{p84}}$} & \colhead{$z_\mathrm{min}$} & \colhead{$z_\mathrm{max}$} & \colhead{Type}\\
\colhead{(1)} & \colhead{(2)} & \colhead{(3)} & \colhead{(4)} & \colhead{(5)} & \colhead{(6)} & \colhead{(7)} & \colhead{(8)}
}
\startdata
587724242842026102 & 0.0139 & 10.36 & 10.28 & 10.41 & 0.0137 & 0.0732 & 3\\
587728669882515508 & 0.0150 & 10.08 & 9.96 & 10.15 & 0.0122 & 0.0658 & 3\\
587729386602692631 & 0.0105 & 10.04 & 9.98 & 10.09 & 0.0096 & 0.0516 & 3\\
587731174379618333 & 0.0128 & 10.21 & 10.15 & 10.24 & 0.0109 & 0.0587 & 3\\
587731872847429772 & 0.0124 & 10.16 & 10.12 & 10.22 & 0.0115 & 0.0620 & 3\\
587731873384562974 & 0.0146 & 10.06 & 10.00 & 10.15 & 0.0132 & 0.0711 & 3\\
587732484369743922 & 0.0074 & 10.23 & 10.14 & 10.30 & 0.0050 & 0.0158 & 2\\
587732591177695368 & 0.0114 & 10.03 & 9.97 & 10.07 & 0.0101 & 0.0540 & 3\\
587735044151509021 & 0.0146 & 10.04 & 9.87 & 10.08 & 0.0144 & 0.0780 & 3\\
587735342651605021 & 0.0068 & 10.45 & 10.35 & 10.50 & 0.0050 & 0.0110 & 3\\
587736940370788440 & 0.0106 & 10.05 & 9.99 & 10.09 & 0.0099 & 0.0534 & 3\\
587736943056519174 & 0.0126 & 10.24 & 10.18 & 10.27 & 0.0124 & 0.0666 & 3\\
587738067813400649 & 0.0149 & 10.09 & 10.04 & 10.13 & 0.0128 & 0.0690 & 3\\
587738196660518935 & 0.0149 & 10.20 & 10.16 & 10.22 & 0.0133 & 0.0724 & 3\\
587738372745593177 & 0.0133 & 10.24 & 10.21 & 10.25 & 0.0113 & 0.0606 & 3\\
587738952033304694 & 0.0133 & 10.18 & 10.12 & 10.21 & 0.0126 & 0.0678 & 3\\
587739114700669144 & 0.0148 & 10.29 & 10.20 & 10.37 & 0.0145 & 0.0773 & 3\\
587739406255783959 & 0.0145 & 10.06 & 10.00 & 10.16 & 0.0131 & 0.0705 & 4\\
587739407328411721 & 0.0145 & 10.29 & 10.21 & 10.34 & 0.0129 & 0.0693 & 3\\
587739407863906479 & 0.0146 & 10.20 & 10.13 & 10.23 & 0.0132 & 0.0710 & 3\\
\enddata
\tablecomments{Columns: (1) SDSS object ID, (2) spectroscopic redshift, (3) total stellar mass, (4) 16th percentile of the total mass PDF, (5) 84th percentile of the total mass PDF, (6) minimum redshift at which the galaxy satisfies our sample selection criteria, (7) maximum redshift at which the galaxy satisfies our sample selection criteria, (8) profile type. (This table is available in its entirety in a machine-readable form in the online journal.  A portion is shown here for guidance regarding its form and content.)\label{sersic_mass_table_nodust}} 
\end{deluxetable*}

\begin{turnpage}
\begin{deluxetable*}{lcrrrrrrrrrrrrrr}
\tablewidth{0pc}
\tablecaption{Total, bulge, and disk stellar mass estimates from bulge+disk photometry with dust-free models}
\tablehead{\colhead{objID} & \colhead{$z$} & \colhead{$\log M$} & \colhead{$\log M_{\mathrm{p16}}$} & \colhead{$\log M_{\mathrm{p84}}$} & \colhead{$\log M_\mathrm{B}$} & \colhead{$\log M_{\mathrm{B},\,\mathrm{p16}}$} & \colhead{$\log M_{\mathrm{B},\,\mathrm{p84}}$} & \colhead{$\log M_\mathrm{D}$} & \colhead{$\log M_{\mathrm{D},\,\mathrm{p16}}$} & \colhead{$\log M_{\mathrm{D},\,\mathrm{p84}}$} & \colhead{$z_\mathrm{min}$} & \colhead{$z_\mathrm{max}$} & \colhead{\pps} & \colhead{Type} & \colhead{$\Delta_\mathrm{B+D}$}\\
\colhead{(1)} & \colhead{(2)} & \colhead{(3)} & \colhead{(4)} & \colhead{(5)} & \colhead{(6)} & \colhead{(7)} & \colhead{(8)} & \colhead{(9)} & \colhead{(10)} & \colhead{(11)} & \colhead{(12)} & \colhead{(13)} & \colhead{(14)} & \colhead{(15)} & \colhead{(16)}
}
\startdata
587731872847429772 & 0.0124 & 10.18 & 10.08 & 10.21 & 9.35 & 9.29 & 9.40 & 10.08 & 9.95 & 10.13 & 0.0115 & 0.0621 & 0.002 & 3 & 0.24\\
587739406255783959 & 0.0145 & 10.06 & 10.00 & 10.17 & 9.94 & 9.86 & 9.99 & 9.85 & 9.72 & 9.91 & 0.0131 & 0.0705 & 1.000 & 4 & 1.40\\
587731174379618333 & 0.0128 & 10.20 & 10.14 & 10.23 & 9.20 & 9.15 & 9.22 & 10.05 & 10.01 & 10.09 & 0.0109 & 0.0589 & 0.000 & 3 & 1.35\\
587729772607045903 & 0.0143 & 10.14 & 9.91 & 10.21 & 9.19 & 9.11 & 9.24 & 9.86 & 9.75 & 10.08 & 0.0128 & 0.0692 & 0.057 & 3 & 0.91\\
587729386602692631 & 0.0105 & 10.00 & 9.94 & 10.06 & 9.69 & 9.66 & 9.72 & 9.73 & 9.62 & 9.78 & 0.0096 & 0.0516 & 0.000 & 3 & 0.14\\
587728880332308694 & 0.0124 & 10.26 & 10.15 & 10.37 & 10.33 & 10.30 & 10.36 & 9.64 & 9.56 & 9.71 & 0.0122 & 0.0657 & 0.018 & 3 & 1.53\\
587725981765664905 & 0.0142 & 10.10 & 10.04 & 10.13 & 8.90 & 8.75 & 8.98 & 10.00 & 9.86 & 10.06 & 0.0108 & 0.0581 & 0.002 & 3 & 0.65\\
587724242842026102 & 0.0139 & 10.17 & 10.05 & 10.25 & 9.86 & 9.83 & 9.86 & 9.96 & 9.77 & 10.02 & 0.0137 & 0.0736 & 0.000 & 3 & 0.34\\
587722981745295552 & 0.0234 & 10.25 & 10.06 & 10.34 & 9.74 & 9.70 & 9.74 & 10.02 & 9.80 & 10.14 & 0.0140 & 0.0754 & 0.004 & 3 & 0.27\\
587722981747392587 & 0.0187 & 10.14 & 10.07 & 10.21 & 8.79 & 8.61 & 8.92 & 10.15 & 10.02 & 10.23 & 0.0153 & 0.0827 & 1.000 & 2 & 0.17\\
587735349648752750 & 0.0230 & 10.28 & 10.06 & 10.41 & 9.79 & 9.76 & 9.81 & 10.23 & 10.09 & 10.32 & 0.0168 & 0.0897 & 0.015 & 3 & 0.45\\
587735349647442032 & 0.0246 & 10.21 & 10.13 & 10.27 & 8.36 & 7.98 & 8.54 & 10.16 & 10.07 & 10.23 & 0.0052 & 0.0290 & 0.964 & 2 & 0.37\\
587735349645934646 & 0.0202 & 10.49 & 10.39 & 10.56 & 10.52 & 10.49 & 10.54 & 9.78 & 9.62 & 9.88 & 0.0184 & 0.0953 & 0.000 & 4 & 1.26\\
587735349636300832 & 0.0185 & 10.24 & 10.03 & 10.32 & 10.20 & 10.16 & 10.23 & 9.65 & 9.53 & 9.72 & 0.0161 & 0.0862 & 0.000 & 3 & 0.50\\
587735349111226443 & 0.0234 & 10.16 & 10.08 & 10.23 & 10.21 & 10.14 & 10.32 & 8.80 & 8.53 & 9.04 & 0.0145 & 0.0775 & 0.509 & 1 & 0.64\\
587736541478519037 & 0.0232 & 10.23 & 10.20 & 10.23 & 9.61 & 9.59 & 9.61 & 9.87 & 9.66 & 9.97 & 0.0117 & 0.0628 & 0.029 & 3 & 1.99\\
587736540944269327 & 0.0239 & 10.67 & 10.58 & 10.75 & 10.83 & 10.80 & 10.85 & 7.73 & 7.54 & 7.96 & 0.0221 & 0.1114 & 0.967 & 1 & 2.18\\
587736540940664928 & 0.0232 & 10.45 & 10.39 & 10.52 & 10.29 & 10.20 & 10.39 & 10.16 & 10.03 & 10.26 & 0.0215 & 0.1092 & 0.000 & 3 & 0.85\\
588023670242410560 & 0.0236 & 10.43 & 10.33 & 10.49 & 9.71 & 9.68 & 9.71 & 10.15 & 10.03 & 10.23 & 0.0102 & 0.0548 & 0.036 & 3 & 1.25\\
587736526442266845 & 0.0231 & 10.01 & 9.87 & 10.20 & 8.77 & 8.60 & 8.90 & 10.00 & 9.85 & 10.18 & 0.0135 & 0.0728 & 1.000 & 2 & 0.07\\
\enddata
\tablecomments{Columns: (1) SDSS object ID, (2) spectroscopic redshift, (3) total stellar mass, (4) 16th percentile of the total mass PDF, (5) 84th percentile of the total mass PDF, (6) bulge stellar mass, (7) 16th percentile of the bulge stellar mass PDF, (8) 84th percentile of the bulge mass PDF, (9) disk stellar mass, (10) 16th percentile of the disk stellar mass PDF, (11) 84th percentile of the disk mass PDF, (12) minimum redshift at which the galaxy satisfies our sample selection criteria, (13) maximum redshift at which the galaxy satisfies our sample selection criteria, (14) $F$-test probability that two components are \emph{not} required, (15) profile type, (16) difference between the total mass and sum of the bulge and disk masses in units of the standard error. (This table is available in its entirety in a machine-readable form in the online journal.  A portion is shown here for guidance regarding its form and content.)\label{bd_mass_table_nodust}} 
\end{deluxetable*}	
\end{turnpage}

\begin{figure}
\centering
\includegraphics[scale=0.85]{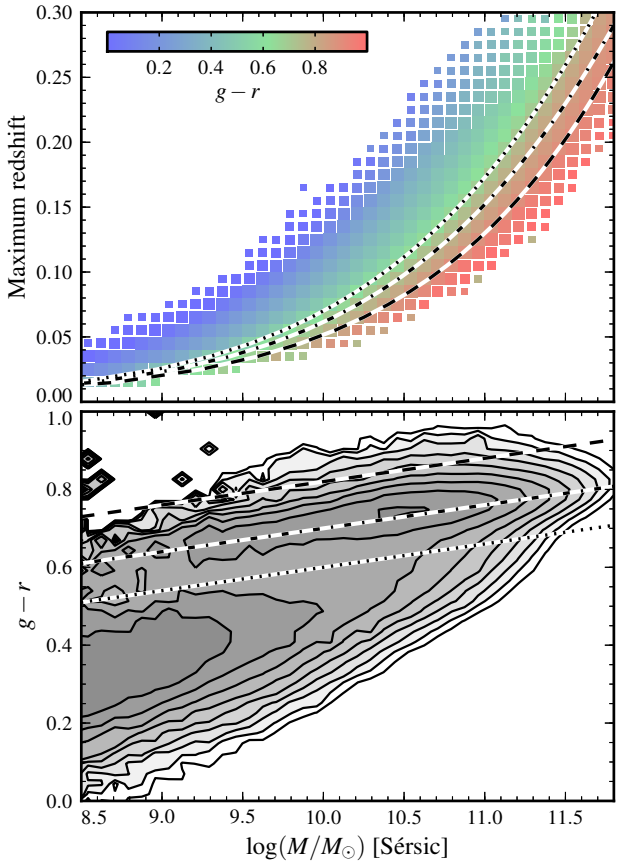}
\caption{Completeness limits as a function of color and stellar mass.  In the top panel we plot the maximum redshift at which galaxies satisfy our sample flux limits, \zmax, as a function of stellar mass.  Tiles are color-coded by the mean \gr color in each bin, and tile size indicates the number of galaxies on a logarithmic scale.  Dashed, dot-dashed and dotted lines show the redshift completeness limits as a function of stellar mass for different fixed-color samples as indicated in the bottom panel and given in Table \ref{mass_lim_params}.  In the bottom panel, contours show the $V_\mathrm{max}$ corrected color--stellar mass relation for all galaxies in our sample.  Contours are logarithmically spaced. Dotted, dot-dashed and dashed lines indicate color as a function of stellar mass for galaxies in the green valley, red sequence, and 3$\sigma$ above the red sequence.}
\label{fig:mass_z}
\end{figure}

\subsection{Mass completeness limits}
\label{completeness}

In order to construct mass-complete samples from the flux-limited SDSS we require some knowledge of galaxies underlying SEDs.  Since our estimates of stellar mass rely on determining the most likely SED given a set of observed $ugriz$ photometry, we can use the best fit model (or the likelihood distribution over a set of models) to estimate the redshift range over which a given galaxy satisfies the flux limits of our sample (i.e. $14 \leq m_r \leq 17.77$).

In Figure \ref{fig:mass_z}, we show the distribution of \zmax---the maximum redshift at which a galaxy still satisfies our sample flux limit---as a function of stellar mass.  In the top panel, we have coded tiles according to galaxies' \gr color using the scale shown in the upper left.  This shows clearly that blue galaxies are observable over a larger volume than red galaxies at fixed stellar mass---or, alternatively, blue galaxies are sampled to lower stellar masses than red in a fixed volume---as one might expect based on SED shape alone.  After some experimentation we found that the redshift-dependent stellar mass limit is well described by a relation of the form

\begin{equation}
\log (M_\mathrm{lim}/M_\odot) = \alpha + \beta\log z + \gamma\log(1+z).
\label{eqn:mass_lim}
\end{equation}

\noindent By fitting the stellar mass--\zmax relation for galaxies with different rest-frame colors, we can provide a more quantitative description of the sampling effects discussed above.  The dotted, dot-dashed and dashed lines in Figure \ref{fig:mass_z} show the resulting functional fits for 3 different color cuts at the green valley, red sequence, and 3$\sigma$ above the red sequence, which have been determined based on the color--stellar mass distribution in the bottom panel of Figure \ref{fig:mass_z}.  The relevant parameters for these fits (as well as several others) are given in Table \ref{mass_lim_params}, and can be used to select mass-complete samples as required.

\begin{deluxetable}{lcccc}
\tablewidth{0pc}
\tablecaption{Best-fit parameters for Equation \ref{eqn:mass_lim}}
\tablehead{\colhead{Galaxy population} & \colhead{\gr\tablenotemark{1}} &\colhead{$\alpha$} & \colhead{$\beta$} & \colhead{$\gamma$}}
\startdata
Green valley	& 0.60				& 12.363	& 2.145  & 3.932\\ 
Red sequence	& 0.70				& 12.575	&  2.203 & 3.696\\ 
Red sequence $+1\sigma$ & 0.74		& 12.476	&  2.108 & 4.845\\ 
Red sequence $+2\sigma$ & 0.78		& 12.791	&  2.273 & 2.932\\ 
Red sequence $+3\sigma$ & 0.82		& 12.994	&  2.353 & 1.734
\enddata
\label{mass_lim_params}
\tablenotetext{1}{Color here refers to galaxy color at $\log (M/M_\odot) = 10$, where the slope of the color mass relation is taken as 0.06 (see Figure \ref{fig:mass_z})}.
\end{deluxetable}

\subsection{When is a bulge not a bulge?}
\label{quality_control}

Up to this point we have proceeded without considering the physical reality of structural components recovered by our bulge+disk decompositions.  While by construction the recovered components represent the best numerical match to the data, for most scientific investigations we require that the recovered components and their properties be \emph{astrophysically} meaningful as well.  In practice, it is useful to consider a given fit's reliability as the combination of two separate questions:  i) to what extent does a given galaxy image support modelling by a more complex photometric model, e.g., two components versus one? and ii) when can we treat multiple fitted components as physically meaningful structures?

In \citetalias{simard2011}, the distinction between single and multi-component decompositions was made using an $F$-test to compare the $\chi^2$ values of their three fitted models (single S\'ersic, de Vaucouleurs bulge+disk, and S\'ersic bulge+disk).  They define a parameter, \pps, as the probability that a multi-component model is \emph{not} required relative to a single component (i.e. S\'ersic profile) fit.  To put it another way, \pps gives the probability that a S\'ersic model is adequate to describe the data, and can therefore be used to distinguish genuine structural components \citep[see also][]{meert2013}.  For this reason, we include the \pps values computed by \citetalias{simard2011} in our bulge+disk mass catalogs.  As shown by \citetalias{simard2011}, the SDSS data are generally only sufficient to distinguish multiple components in galaxies clearly hosting both bulges and disks ($0.2 \leq B/T \leq 0.45$), and adopting even modest cuts on the $F$-test probability, e.g. $P_\mathrm{pS} \leq 0.32$, seems to limit the incidence of spurious bulges and disk based on, for example, the distribution of disk axial ratios.  As discussed by \citet{meert2013}, the $F$-test can be used to select relatively pure samples of two-component galaxies, but these samples are not necessarily complete.  Some caution must therefore be used in blindly applying \pps to select bulge and disk subsamples.

Alternatively, one can attempt to classify the fitted profiles in terms of their properties, singling out galaxies where two distinct components appear to be physically motived and discarding cases where multiple components are either unjustified by the data, as above, or nonsensical.  Both \citet{allen2006} and \citet[hereafter \citetalias{lackner2012}]{lackner2012} have employed such an approach in their analyses of galaxy structure, and we follow them in filtering the \citetalias{simard2011} data.  In their work, \citet{allen2006} identify eight profile types that encompass the range of decompositions observed in their sample of $\sim$10,000 galaxies.  These can be broadly understood in terms of the physical systems they describe: two profile types describe single-component models, either pure bulge or disk, while the remaining six describe variations of two-component bulge+disk systems.  \citet{allen2006} describe the application of a logical filter to classify decompositions as reliable, in which case parameters of the bulge+disk fits are retained, or unreliable, in which case structural parameters of single-component fits are adopted.  In classifying the \citetalias{simard2011} data we adopt a simplified framework in which we divide the galaxy population into only 4 classes based on their best-fiting {\sc gim2d} profiles.  This scheme is summarized below.\\

\begin{figure*}
\centering
\includegraphics[scale=0.90]{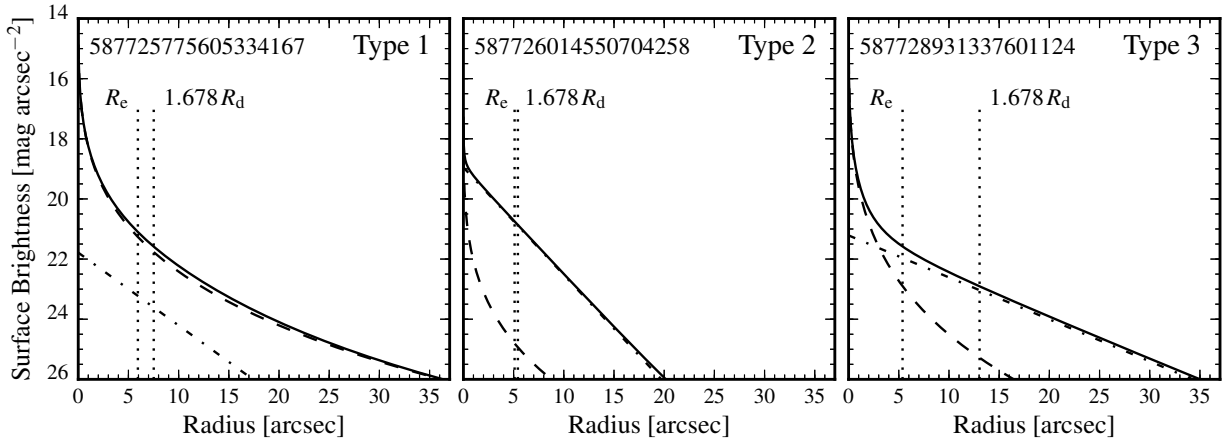}
\caption{Examples of profile types 1, 2, and 3 described in Section \ref{quality_control}.  Solid lines show total surface brightness profiles derived from the bulge+disk fits, while dashed and dot-dashed lines show the surface brightness profiles of the bulge and disk separately.  Dotted lines mark the bulge and disk half-light size.}
\label{fig:profile_type}
\end{figure*}

\begin{figure}
\centering
\includegraphics[scale=0.90]{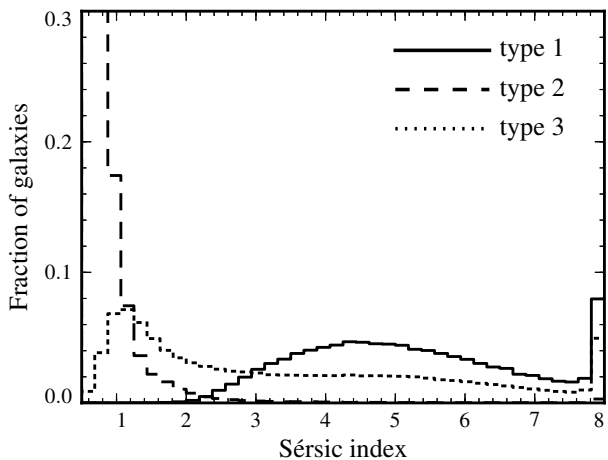}
\caption{Distribution of S\'ersic indices for the different profile types described in Section \ref{quality_control}.  Type 1, 2, and 3 profiles are classified as elliptical, disk, and bulge+disk based on the relative significance (in terms of surface brightness) of the structural components measured by {\sc gim2d}.}
\label{fig:sersic_type}
\end{figure}

\hangindent=0.5cm
\hangafter=0
\noindent (i) \emph{Type~1} A de Vaucouleurs component dominates the surface brightness profile at all radii, or no exponential component is fit (i.e. $B/T = 1$).  We also include here galaxies for which the best-fit disk scale length is less than 0.2 pixels---i.e. the disk is collapsed to a point source---regardless of $B/T$ ($\sim$0.5\% of fits).  In terms of the single component S\'ersic profiles fit by \citetalias{simard2011}, the distribution of $n$ for these profiles peaks at $n\sim$~4-5.  These are single component elliptical galaxies, and are related to profile types 5 and 8 from \citet{allen2006}.\\
\\
(ii) \emph{Type~2} The exponential component dominates the surface brightness profile at all radii, or no de Vaucouleurs component is fit (i.e. B/T = 0).  As for type 1 profiles, we include here any fits where the half-light radius of the best-fit de Vaucouleurs profile is less than 0.2 pixels ($\sim$2.6\% of fits).  These galaxies have low $n$ in the single S\'ersic fits, and are either pure disk systems or disks hosting a weak central pseudo-bulge.  These galaxies are related to profile types 2, 5, 7, and 8 from \citet{allen2006}.\\
\\
(iii) \emph{Type~3} The de Vaucouleurs component dominates the surface brightness profile in the central regions, while the exponential component dominates at large radii.  The de Vaucouleurs and exponential profiles cross only once at an $r$-band surface brightness, $\mu_r$, less than 26 mag arcsec$^{-2}$.  For the most part these are genuine bulge+disk systems, however in some cases these are single-component (elliptical) galaxies in which a disk has been included to account for deviations from a pure de Vaucouleurs profile, e.g., tidal features, isophotal twists, or $n \gtrsim 4$.  Such spurious fits stand out in the distribution of axis ratios as an excess of face-on disks \citep[e.g. figure 13 of \citetalias{simard2011};][\citetalias{lackner2012}]{allen2006,benson2007}.  These are classified as type 1 profiles by \citet{allen2006}.\\
\\
(iv) {\it Type~4} These galaxies include everything that cannot be classified as types 1, 2, or 3.  These include galaxies where the bulge and disk surface brightness profiles cross twice at $\mu_r < 26$ mag arcsec$^{-2}$ , or where the disk and bulge profiles are inverted---i.e. the disk profile dominates in the central regions and the bulge dominates at large radii.  These encompass profile types 3, 4 and 6 from \citet{allen2006}. \\

\noindent  Examples of the first three profile types (i.e. elliptical, disk, and bulge+disk) are shown in Figure \ref{fig:profile_type}.  In our sample, 12.4\% of galaxies are classified as type 1, 11.7\% as type 2, 71.7\% as type 3, and 4.2\% as type 4 profiles.  In Figure \ref{fig:sersic_type} we show the distribution of S\'ersic indices for profile types 1, 2, and 3.  The classification of type 1 and 2 profiles as ellipticals and disks respectively is supported by the distribution of $n$ for these galaxies.  While the majority of galaxies are classified as type 3 (bulge+disk) profiles, relatively few of these seem to \emph{require} a multi-component model to describe their surface-brightness profiles; only 37\% of type 3 profiles satisfy the cut of $P_\mathrm{pS} \leq 0.32$ discussed above.  The broad range of galaxies classified as type 3 is also evident in their S\'ersic index distribution.  In most instances, it is therefore useful (or required!) to combine the profile type with cuts on the $F$-test probability when selecting bulge and/or disk subsamples.

As a test of our classifications, we can compare the profile types determined here with those assigned by \citetalias{lackner2012} in their analysis of SDSS galaxies.  Briefly, \citetalias{lackner2012} fit a sample of 71,825 galaxies with 5 different image models: two multi-component (bulge+disk) and three single component profiles.  In order to determine the most appropriate model for each galaxy, they use a combination of physically-driven quality assessment and statistically-motivated goodness of fit.  Their final classification is in to one of 5 categories: de Vaucouleurs bulge plus disk, pseudo-bulge ($n=1$) plus disk, pure de Vaucouleurs profile, pure exponential profile, or S\'ersic profile.  In total there are 66,693 galaxies in common between our sample and that of \citetalias{lackner2012}.  It is worth noting that the classifications of \citetalias{lackner2012} incorporate additional information about the likelihood of a bulge+disk fit relative to other fitted profiles, similar to the \pps parameter described above.  We have chosen to leave these as two separate diagnostics in order to allow greater flexibility in how samples are selected from within our catalog.  Nevertheless, a comparison of the raw profile classifications, independent of the $F$-test probabilities, is still useful in understanding the behavior of these different diagnostics.

Of type 1 galaxies, 44\% are classified as ellipticals by \citetalias{lackner2012}, with an additional 21\% and 28\% classified as de Vaucouleurs bulge+disk and single S\'ersic profiles, respectively.  Despite the relatively poor agreement between the raw classifications for type 1 profiles, the distribution of S\'ersic indices determined by \citetalias{lackner2012} for these galaxies is clearly peaked around $n \approx 4$ (median value of 3.9), supporting our classification of these systems as ellipticals.  The agreement between our pure disk classifications is significantly better, with 66\% of type 2 galaxies classified as pure exponential disks by \citetalias{lackner2012} and a further 32\% classified as single S\'ersic profiles.  As for type 1 profiles, the distribution of S\'ersic indices for type 2 galaxies seems to support their classification as disks (median $n \approx 0.84$).  In total, 24\% of galaxies are classified as being best fitted by either a pure de Vaucouleurs or pure exponential disk model.

There is less clear agreement between our type 3 (bulge+disk) galaxies and the classifications of \citetalias{lackner2012}; only 36\% of type 3 galaxies are classified as hosting two-components by \citetalias{lackner2012}, with the majority of remaining galaxies classified as unknown (i.e, fitted with S\'ersic profiles).  As discussed above, our raw classifications make no account for the degree to which a two-component is required by the data, which the classifications of \citetalias{lackner2012} do.  If we adopt a cut of $P_\mathrm{pS} \leq 0.32$---i.e. require that a two-component models is preferred at the $\sim$1$\sigma$ level---then 51\% of the remaining type 3 galaxies are classified as bulge+disk systems by \citetalias{lackner2012} (38\% de Vaucouleurs bulge+disk and 13\% pseudo-bulge+disk).  As might be expected, increasing the required significance of the two-component fit improves agreement between the two classification schemes, but at the expense of the total number of bulge+disk galaxies.  Overall, the comparison with \citetalias{lackner2012} supports running theme of this section: identifying genuine bulge+disk systems requires a combination of quality assessment metrics which incorporate information about both the physical and numerical likelihood of a given decomposition.

 \subsection{Identifying spurious bulge and disk data}

Although we have taken steps to limit the number of galaxies with suspect mass estimates, the final catalogs will inevitably contain values that, for one reason or another, are unreliable.  As discussed in Section \ref{mass_comp}, one way to identify such spurious values is to identify data which are internally inconsistent, e.g. when $M_\mathrm{B} + M_\mathrm{D}$ significantly exceeds $M_\mathrm{B+D}$.  In Tables \ref{bd_mass_table} and \ref{bd_mass_table_nodust} we include an additional parameter, $\Delta_\mathrm{B+D}$, which gives the offset between $M_\mathrm{B} + M_\mathrm{D}$ and $M_\mathrm{B+D}$ in units of the standard error.  Galaxies for which the bulge and disk masses exceed the total mass by more than 1$\sigma$ should be approached with extreme caution, and we recommend in these cases using the dust-free masses as more reliable estimate for the ``true'' stellar mass.

\section{Summary}

In this paper we present a catalog of stellar mass estimates for 657,247 galaxies selected from the SDSS DR7.  The foundation for these mass estimates is the reprocessed SDSS $g$- and $r$-band photometry published by \citet{simard2011}, which we have now extended to include the $u$, $i$, and $z$ bands.  In addition to incorporating improved sky background estimates and object deblending, the \citet{simard2011} catalog provides structural information for the best-fit single-component S\'ersic and bulge+disk photometric models on a galaxy-by-galaxy basis, which we use to compute masses separately for bulges and disks.

Masses are derived by comparing galaxies' photometry to a grid of synthetic models which span a range of possible ages, star-formation histories, metallicities, and dust extinctions.  We have paid careful attention when conducting these SED fits to account for non-detections and upper limits, which require a modified treatment of the likelihood function.  In the end, statistical uncertainties on individual mass estimates are $\sim$0.15 dex ($\sim$40\%).

We have shown that additional systematic uncertainties \emph{must} be accounted for when considering photometrically-derived stellar masses.  Changing between SPS models can lead shifts of 0.1 to 0.2 dex depending on the age of the stellar population, and changes in the IMF and extinction law contribute at a similar level.  Particular care must be taken when changing the slope of the IMF at high masses, as this can lead to \emph{differential} changes in the mass of  star-forming and passive galaxies at the level of 0.1 dex. From a stellar evolutionary standpoint, we have explored the particular influence of BHB, BS, and tp-AGB stars on the derived masses.  In all cases, systematic uncertainties are below $\pm$0.1 dex, with the largest effects arising from changes in the temperature and luminosity of tp-AGB stars.  These systematic effects are most important in young stellar populations, where AGB stars contribute significantly to the overall flux. 

Full catalogs containing total, bulge and disk stellar mass estimates for both dusty and dust-free models are included here, along with additional parameters to aid in identifying robust subsamples. We stress that, while we provide estimates of bulge and disk mass for all galaxies on the basis of their bulge+disk decompositions, particular care should be taken to select components that are both numerically and scientifically justified.

\acknowledgements

SLE and DRP gratefully acknowledge the receipt of NSERC Discovery Grands which funded this research.  We are grateful to the anonymous referee, whose comments helped to improve the manuscript, and thank Asa Bluck and Genevieve Graves for useful discussions.

Funding for the SDSS and SDSS-II has been provided by the Alfred P. Sloan Foundation, the Participating Institutions, the National Science Foundation, the U.S. Department of Energy, the National Aeronautics and Space Administration, the Japanese Monbukagakusho, the Max Planck Society, and the Higher Education Funding Council for England. The SDSS Web Site is http://www.sdss.org/.

The SDSS is managed by the Astrophysical Research Consortium for the Participating Institutions. The Participating Institutions are the American Museum of Natural History, Astrophysical Institute Potsdam, University of Basel, University of Cambridge, Case Western Reserve University, University of Chicago, Drexel University, Fermilab, the Institute for Advanced Study, the Japan Participation Group, Johns Hopkins University, the Joint Institute for Nuclear Astrophysics, the Kavli Institute for Particle Astrophysics and Cosmology, the Korean Scientist Group, the Chinese Academy of Sciences (LAMOST), Los Alamos National Laboratory, the Max-Planck-Institute for Astronomy (MPIA), the Max-Planck-Institute for Astrophysics (MPA), New Mexico State University, Ohio State University, University of Pittsburgh, University of Portsmouth, Princeton University, the United States Naval Observatory, and the University of Washington. 

\bibliographystyle{apj}
\bibliography{biblio}

\twocolumngrid

\begin{appendix}

\section{Comparison of repeat $r$-band S\'ersic profile measurements}
\label{fit_comp}

As discussed in Section \ref{data}, each of the $u$, $i$, and $z$ bands were fitted simultaneously with the $r$-band using structural parameters---bulge and disk size, position angle, and ellipticity/inclination---taken from the joint \emph{gr} fits published by \citetalias{simard2011}.  In the case of bulge+disk model fits, $B/T$ and total magnitude vary independently in each band, and therefore measurements of the $r$-band bulge, disk and total properties are nearly identical between different band combinations (i.e. \emph{ur}, \emph{gr}, \emph{ir}, or \emph{zr}).  In the case of our S\'ersic profile fits, however, the S\'ersic index was fitted independently for each \emph{pair} of bands, and therefore the derived $r$-band magnitude and best-fit S\'ersic index depend on the particular band it is coupled to.  In our mass estimates we adopt the $r$-band values given by \citetalias{simard2011} using the combined \emph{g}- and \emph{r}-band data.  Comparing the derived $r$-band magnitude between the different band pairs provides some estimate of the uncertainty induced by this choice.

\begin{figure}
\centering
\includegraphics[scale=0.90]{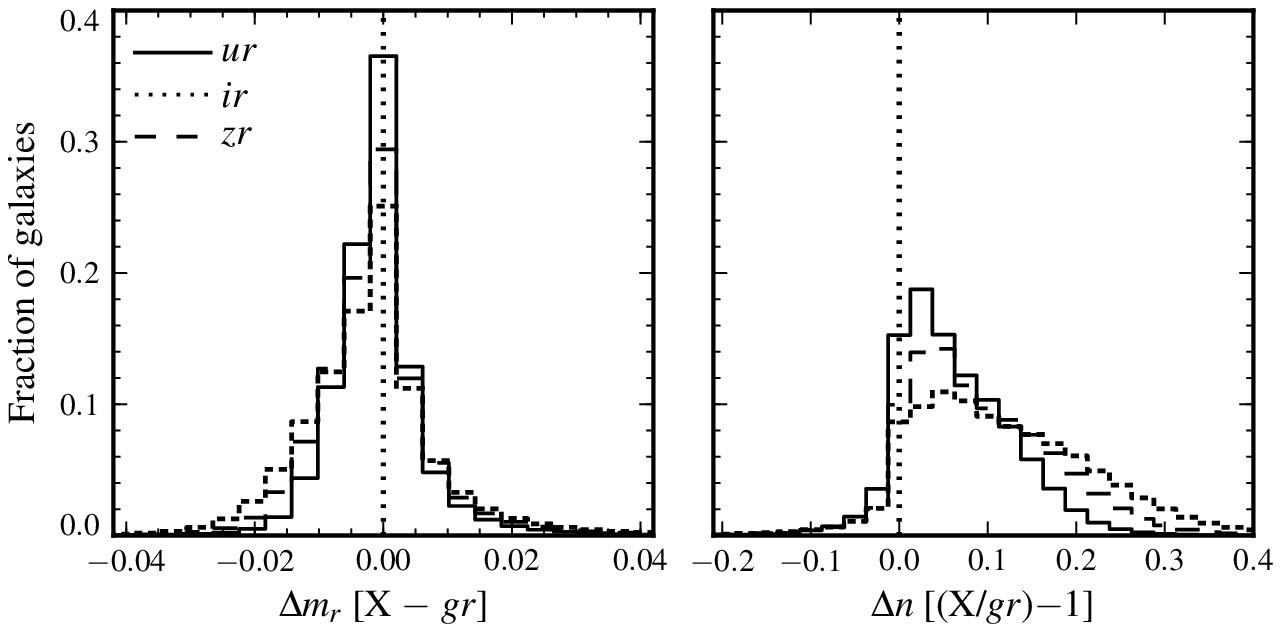}
\caption{Comparison of repeat measurements of the $r$-band apparent magnitude (left panel) and S\'ersic index (right panel) discussed in Appendix \ref{fit_comp}.  Different histograms show the offset of the \emph{ur}, \emph{ir}, and \emph{zr} fits from the \emph{gr} values published by \citetalias{simard2011}.  The dotted vertical line indicates no offset, and is shown for reference only.}
\label{fig:sersic_comp}
\end{figure}

In Figure \ref{fig:sersic_comp} we show a comparison of the $r$-band magnitude, $m_r$, and S\'ersic index, $n$, for each of the \emph{ur}, \emph{ir}, and \emph{zr} S\'ersic fits.  In each panel we plot the difference relative to the \emph{gr} values published by \citetalias{simard2011}.  We find that the pairwise fits yield estimates of $m_r$ that agree to better than a few percent regardless of band pair.  This is less than the typical photometric uncertainty on an individual $m_r$ measurement, and therefore we don't expect that the derived mass should depend significantly on our adoption of measurements from the \emph{gr} S\'ersic fits.

The derived S\'ersic index shows more significant variability from band to band, in agreement with previous work \citep[e.g.][]{kelvin2012,vika2013}.  The scatter between the \emph{gr} fits and other bands is 10-20\% on average, but differences can be up to 40\% in extreme cases.  The fits also show a tendency towards systematically higher $n$ by 5, 10, and 8\% in the \emph{ur}, \emph{ir}, and \emph{zr} fits.  

In the case of bulge+disk fits, while structural parameters are fixed across bands, the $B/T$ is determined independently.  This allows the bulge+disk fits more freedom to accommodate global trends in, for example, color as a function of radius.  The same is not true of the S\'ersic profile fits; this ``stiffness'' of the S\'ersic models is reflected in the comparison of $m_r$ and $n$ for the S\'ersic fits, where the largest differences are observed between the \emph{gr} and \emph{ir} fits, i.e. where the individual bands both have relatively high \emph{S/N}.  To some extent these effects can be mitigated by fitting each band individually \citep[e.g.][]{kelvin2012}, albeit at the expense of the \emph{S/N} boost offered by pairwise fits, or by adopting a functional form for the dependence of structural parameters with wavelength \citep[e.g.][]{hausler2013}.

\section{The reliability of flux measurements from two-dimensional parametric decompositions}
\label{flux_sims}

Here we investigate the overall reliability of the \citetalias{simard2011} flux estimates used in our computation of stellar masses.  We follow \citet{simard2002} and \citet{simard2009} in addressing this question by means of Monte Carlo simulations, whereby we inject model galaxy images into randomly-selected SDSS imaging frames and re-process them following the procedures outlined in Section \ref{data} and \citetalias{simard2011}.  Our goal is to assess any systematic effects in the parametric recovery of galaxies' flux that may affect our determination of galaxies' stellar mass, as well as statistical uncertainties due to sky background estimation and/or crowding.  In the following, we limit our analysis to the $g$ and $r$ bands; however, we expect that results are similar for the joint $ur$, $ri$, and $rz$ fits.

\subsection{Galaxy simulations}
\label{sim_description}

We generate a large number of smooth galaxy image models---either a single-component S\'ersic profile or a two-component de Vaucouleurs bulge plus disk profile---with a range of structural parameters described in Table \ref{sim_param}.  Synthetic galaxy images are generated using the {\sc gsimul} task included as part of {\sc gim2d}.  In each case, parameters are randomly drawn with uniform probability from their allowed ranges, and the resulting image model is injected into a randomly-selected SDSS corrected frame using the {\sc sextractor} segmentation image as a guide.  We require that the model center falls on a sky pixel, however place no such constraint on the simulated galaxies' isophotal area (see Section \ref{crowding} for a discussion of crowding effects).

The methodology described by \citetalias{simard2011} relies on simultaneously fitting the $g$- and $r$-band images, and our Monte Carlo procedure therefore requires that we generate coupled pairs of $g$- and $r$-band models.  In the case of single-component galaxies generation of these coupled pairs is relatively straight forward: given a randomly-generated $r$-band magnitude, we generate $g$-band magnitudes such that $0.3 \leq g\!-\!r \leq 1.1$.  Other parameters---i.e. size, S\'ersic index, ellipticity and position angle---are drawn uniformly from their respective ranges given in Table \ref{sim_param}.  In the case of bulge+disk models the procedure is somewhat more complicated.  For each (randomly-generated) $r$-band magnitude and \btr, we use empirical relationships between \btr, \btg, and \gr color (given in Table \ref{empirical_relationship}) to generate the corresponding $g$-band magnitude and \btg.  The typical scatter about the empirical \btr versus color and \btr versus \btg relationships are 0.16 mag and 0.08, respectively, and are incorporated in our generation of \btg and color from \btr.  Bulge and disk size, ellipticity/inclination, and position angle are generated uniformly over the ranges specified in Table \ref{sim_param}.  For simplicity we require that the bulge and disk position angles are the same. 

Simulations can only be used to understand the reliability of the \citetalias{simard2011} flux measurements inasmuch as they reflect the properties of real galaxies, and we therefore impose two additional criteria on our simulations.  First, the ability of {\sc gim2d} to accurately recover flux depends sensitively on surface brightness, so that flux estimates of low surface-brightness galaxies or structural components are significantly uncertain.  We therefore require that simulations have total half-light radii less than 15 pixels in order to limit any bias incurred as a result of considering large, low surface-brightness galaxies (for reference, $\sim$91\% of observed galaxies satisfy our adopted size limit).  Second, in observational samples there is significant covariance between bulge and disk size; however, in generating our models we have explicitly decoupled the properties of bulges and disks.  Although we do not want to impose \emph{a priori} a strong relationship between bulge and disk sizes which may bias our results, we do impose a cut such that $r_\mathrm{e} \leq 1.67r_\mathrm{d}$.  We generate a total of 46,000 simulations matching these criteria: 23,000 each with either single- or two-component light profiles.

Finally, it is worth stressing that these simulations are intended to sample the parameter space that galaxies may occupy (in terms of the parameters listed in Table \ref{sim_param}) without any \emph{a priori} assumptions about how real galaxies actually populate this space.

\begin{deluxetable}{lcc}
\tablewidth{0pc}
\tablecaption{Simulated galaxy model parameters}
\tablehead{\colhead{Parameter} & \colhead{Description} & \colhead{Range of values}}
\startdata
$m_r$	& $r$-band apparent magnitude	& $14 \leq m_r \leq 17.7$ \\ 
$(B/T)_r$	& $r$-band bulge-to-total	&  $0 \leq (B/T)_r \leq 1$ \\ 
$n$ & S\'ersic index & $0.5 \leq n \leq 8$ \\
$r_\mathrm{e}$ & bulge half-light size & $0 \leq r_\mathrm{e} \leq 20$ pixels \\
$r_\mathrm{d}$ & disk scale length & $0 \leq 1.67\,r_\mathrm{d} \leq 20$ pixels \\  
$i$	& disk inclination & $ 0.087 \leq \cos i \leq 1$ \\
$e$ & bulge ellipticity & $0 \leq e \leq 0.7$ \\
$\phi_\mathrm{b,d}$ & bulge and disk position angle & $0^{\circ} \leq \phi_\mathrm{b,d} \leq 180^{\circ}$
\enddata
\tablecomments{Parameters for the randomly-generated galaxy image models described in Appendix \ref{flux_sims}. \label{sim_param}} 
\end{deluxetable}

\begin{deluxetable}{lcccc}
\tablewidth{0pc}
\tablecaption{Relationship between \btr, \btg, and \gr}
\tablehead{\colhead{\btr} & \colhead{\btg} &\colhead{$\sigma$(\btg)} & \colhead{\gr} & \colhead{$\sigma$(\gr)}}
\startdata
0.0 & 0.00 & 0.07 & 0.54 & 0.23 \\  
0.1 & 0.09 & 0.08 & 0.62 & 0.19 \\ 
0.2 & 0.18 & 0.08 & 0.71 & 0.18 \\ 
0.3 & 0.29 & 0.08 & 0.78 & 0.16 \\ 
0.4 & 0.41 & 0.07 & 0.84 & 0.15 \\ 
0.5 & 0.51 & 0.07 & 0.87 & 0.15 \\ 
0.6 & 0.61 & 0.07 & 0.88 & 0.13 \\ 
0.7 & 0.71 & 0.09 & 0.87 & 0.14 \\ 
0.8 & 0.82 & 0.10 & 0.86 & 0.14 \\ 
0.9 & 0.92 & 0.12 & 0.87 & 0.15 \\ 
1.0 & 0.99 & 0.11 & 0.90 & 0.18
\enddata
\tablecomments{Empirical relationships used to construct the image pairs described in Appendix \ref{flux_sims}.\label{empirical_relationship}}
\end{deluxetable}

\subsection{Recovered flux}
\label{flux_recovery}

\subsubsection{The reliability of total flux measurements}

\begin{figure*}
\centering
\includegraphics[scale=0.85]{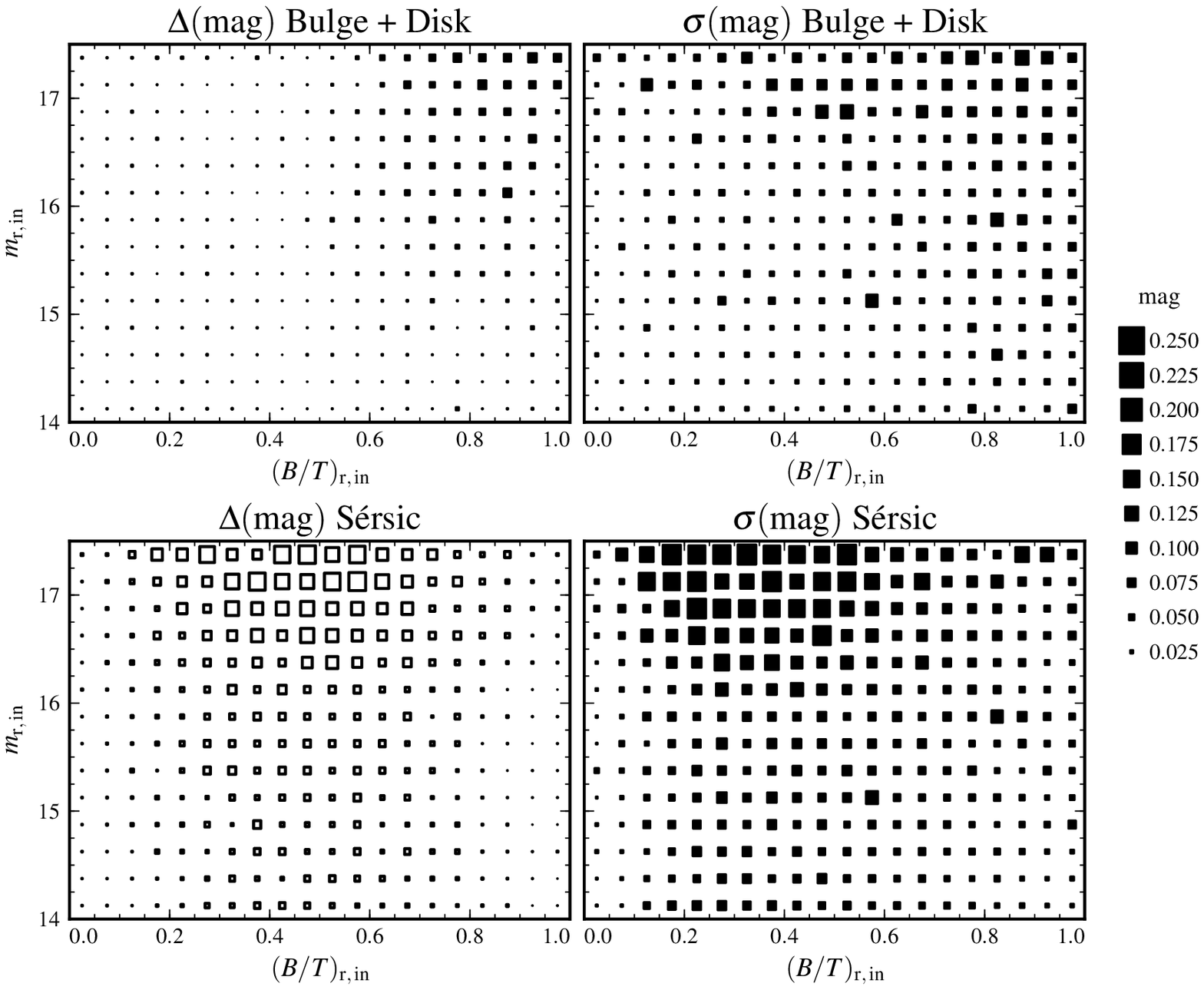}
\caption{Reliability of flux measurements for simulated bulge+disk models.  Systematic and statistical uncertainties are shown as a function of input $B/T$ and apparent magnitude for the Monte Carlo simulations described in Section \ref{sim_description}.  The top and bottom panels show the results of using either a bulge+disk or S\'ersic model to fit the simulated images as indicated.  Left panels show the median systematic offset between the input (simulated) flux and the flux recovered by {\sc gim2d}.  Tiles are scaled according to the legend on the right, while open and filled squares correspond to negative and positive offsets, respectively.  The right panel shows r.m.s.~scatter in recovered magnitudes for the same galaxies.}
\label{fig:total_flux_bd}
\end{figure*}

\begin{figure*}
\centering
\includegraphics[scale=0.85]{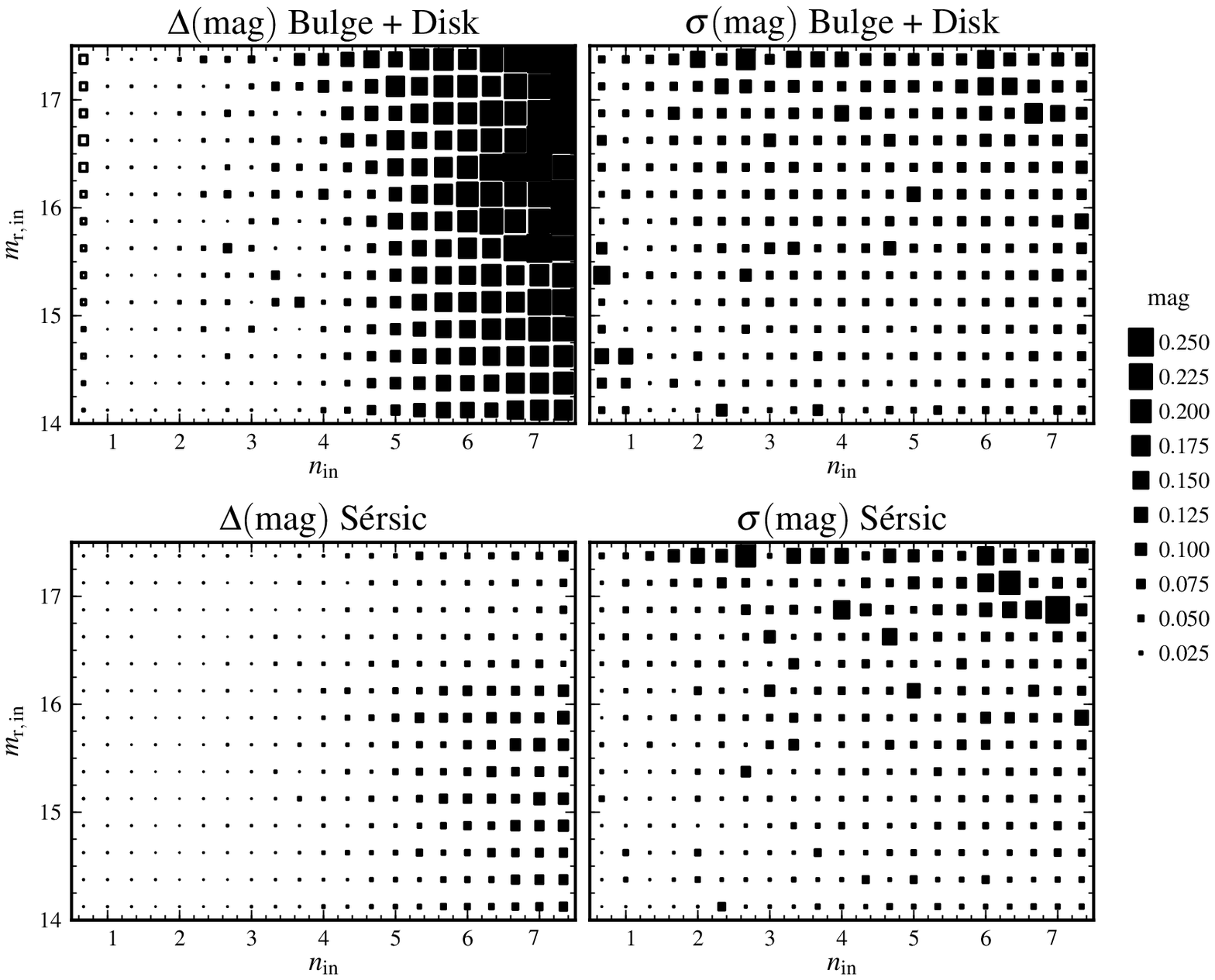}
\caption{Same as Figure \ref{fig:total_flux_bd}, but for single-component input models.  Systematic and statistical uncertainties are shown as a function of input S\'ersic index and apparent magnitude for the Monte Carlo simulations described in Section \ref{sim_description}.  The top and bottom panels show the results of using either a bulge+disk or S\'ersic model to fit the simulated images as indicated.  Left panels show the median systematic offset between the input (simulated) flux and the flux recovered by {\sc gim2d}.  Tiles are scaled according to the legend on the right, while open and filled squares correspond to negative and positive offsets, respectively.  The right panel shows r.m.s.~scatter in recovered magnitudes for the same galaxies.} 
\label{fig:total_flux_sersic}
\end{figure*}

We first consider the recovery of galaxies' total flux, which is summarised in Figures \ref{fig:total_flux_bd} and \ref{fig:total_flux_sersic} for bulge+disk and single-component input models, respectively.  For each simulated image we consider the result of fitting with either a two- or single-component photometric model, indicated as either ``Bulge + Disk'' or ``S\'ersic''.  In each Figure, left panels show the systematic uncertainty in the recovered magnitudes ($m_\mathrm{recovered} - m_\mathrm{input}$) for a range of apparent magnitudes and structural parameters (either $B/T$ or $n$), while the right panels show the statistical uncertainty (i.e.~scatter) in offset values.  Open and filled tiles indicate negative and positive values, and the size of each tile indicates the magnitude of the offset following the scale on the right.  For example, a large filled square represents a significant underestimate of the true flux.

Focusing first on two-component galaxies (Figure \ref{fig:total_flux_bd}), both photometric models produce reasonable estimates of galaxies' total flux for galaxies with $m_r \lesssim 16$ regardless of bulge fraction, which typical systematic offsets $<$0.1 mag.  We stress that these error estimates incorporate both the formal uncertainties output by {\sc gim2d} as well as uncertainties in the sky background.  Pushing towards fainter magnitudes, there is a tendency for S\'ersic profile fits to systematically overestimate the true flux by up to $\sim$0.2 mag, particularly in galaxies with intermediate $B/T$; this overestimate is also reflected in the increased scatter in recovered flux (bottom right panel of Figure \ref{fig:total_flux_bd}.  This primarily reflects the difficulty in constraining the wings of the S\'ersic profile when the absolute sky level is uncertain, particularly at low surface brightness.  In contrast, bulge+disk fits are extremely well behaved over the range of simulated properties, suggesting that they are able to reliably recover the total flux of ``classical'' bulge+disk galaxies. 

Turning to single-component galaxies, the upper left panel in Figure \ref{fig:total_flux_sersic} highlights a particular shortcoming of the bulge+disk modelling adopted here: the combination of a de Vaucouleurs bulge exponential disk profile is unable to accurately recover flux in galaxies with $n \gtrsim 5$.  In this case, the systematic underestimation of flux at high $n$ is governed by an inability of the bulge+disk model to simultaneously reproduce both the bright inner and extended outer profiles of the simulated galaxies, and at least in part motivates the more careful consideration of ``genuine'' bulges and disks using the metrics discussed in Section \ref{quality_control}.  By comparison, the fluxes recovered by S\'ersic profile fits appear to be extremely robust when applied to either simulated sample: systematic uncertainties are less than 0.15 mag across a broad range of input galaxy parameters, and $\sim$0.02 mag for the sample as a whole (i.e. irrespective of $m_r$, $B/T$, or $n$).  

In cases where the only requirement is for robust and reliable total mass measurements, we suggest that masses based on S\'ersic profile fits are to be preferred over bulge+disk models.

\subsubsection{Recovering bulge and disk flux}

\begin{figure*}
\centering
\includegraphics[scale=0.85]{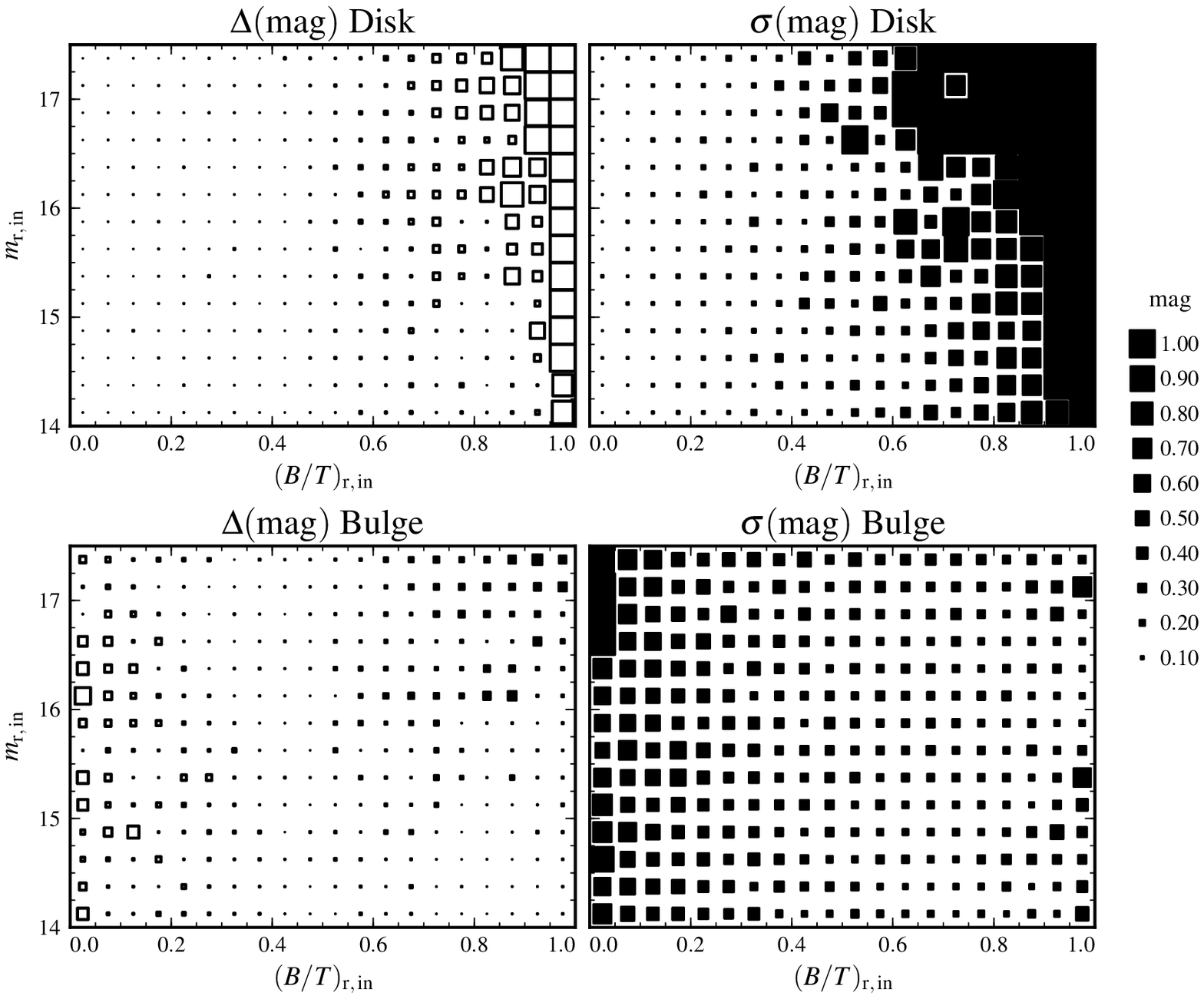}
\caption{Reliability of disk (top panels) and bulge (bottom panels) flux measurements for simulated bulge+disk models.  Systematic and statistical uncertainties are shown as a function of $B/T$ and total apparent magnitude as in Figure \ref{fig:total_flux_bd}.  Tiles are scaled according to either the median offset between input and recovered flux (left panels) or the r.m.s.~scatter in recovered flux (right panels).  Symbols and colors are the same as Figure \ref{fig:total_flux_bd}, however note the factor of four increase in scaling (shown right).}
\label{fig:bulge_disk}
\end{figure*}

As described in Section \ref{sim_description}, our image models are generated from a combination of bulge and disk components.  Therefore, in addition to quantifying the reliability of our overall flux measurements, as above, we can also study separately the recovery of bulge and disk fluxes, which are shown in Figure \ref{fig:bulge_disk}.  The format of this figure is the same as for Figure \ref{fig:total_flux_bd}, except that the size of each tile now reflects the systematic or statistical uncertainty in the recovered disk or bulge (top and bottom panels, respectively) separately---note also the increased scale on the righthand side of Figure \ref{fig:bulge_disk}.  These figures serve to make the point that the ability to reliably recover the flux of an individual structural sub-component is strongly dependent on the galaxies' overall structure, particularly at faint galaxy magnitudes; the recovered bulge (disk) fluxes in galaxies that are disk (bulge) dominated is uncertain by more than 1 mag, while the statistical uncertainty is $\sim$0.5 mag for galaxies with intermediate \btr.  While the precision of bulge and disk fluxes suffer relative to the total flux measurements, they remain relatively accurate in the sense that the systematic uncertainties are generally 0.3--0.4 mag or less, with the exception of disks in bulge-dominated galaxies, whose flux can be significantly overestimated.  For galaxies which host both bulge and disk components---e.g. $0.2 \leq (B/T)_r \leq 0.8$---the typical systematic uncertainty is of order 0.3 mag or less, while statistical uncertainties range from 0.3 to 0.5 mag depending on bulge fraction.

\subsection{Crowding effects}
\label{crowding}

As discussed in the Introduction, one of the principal motivations for the re-derivation of photometric quantities by \citetalias{simard2011} is the relatively poor performance of SDSS DR7 photometry in crowded environments.  Given this, it is worthwhile to assess the performance of {\sc sextractor} and {\sc gim2d} for galaxies with near neighbors.

As described previously, our simulated galaxy images are injected into the SDSS corrected frames requiring only that the model center falls on a ``sky'' pixel (as classified by {\sc sextractor}); therefore, our simulations include instances where the model galaxy's isophotal area overlaps significantly with another photometric object in the image frame.  In order to assess the influence of these near neighbors on the recovered flux, we first find for each simulated image the closest (observed) photometric object with an extinction-corrected $r$-band Petrosian magnitude no more than 2.5 magnitudes fainter than the input model galaxy (i.e., we require that $m_r \leq m_{r,\,\mathrm{input}}+2.5$).  

\begin{figure}
\centering
\includegraphics[scale=0.85]{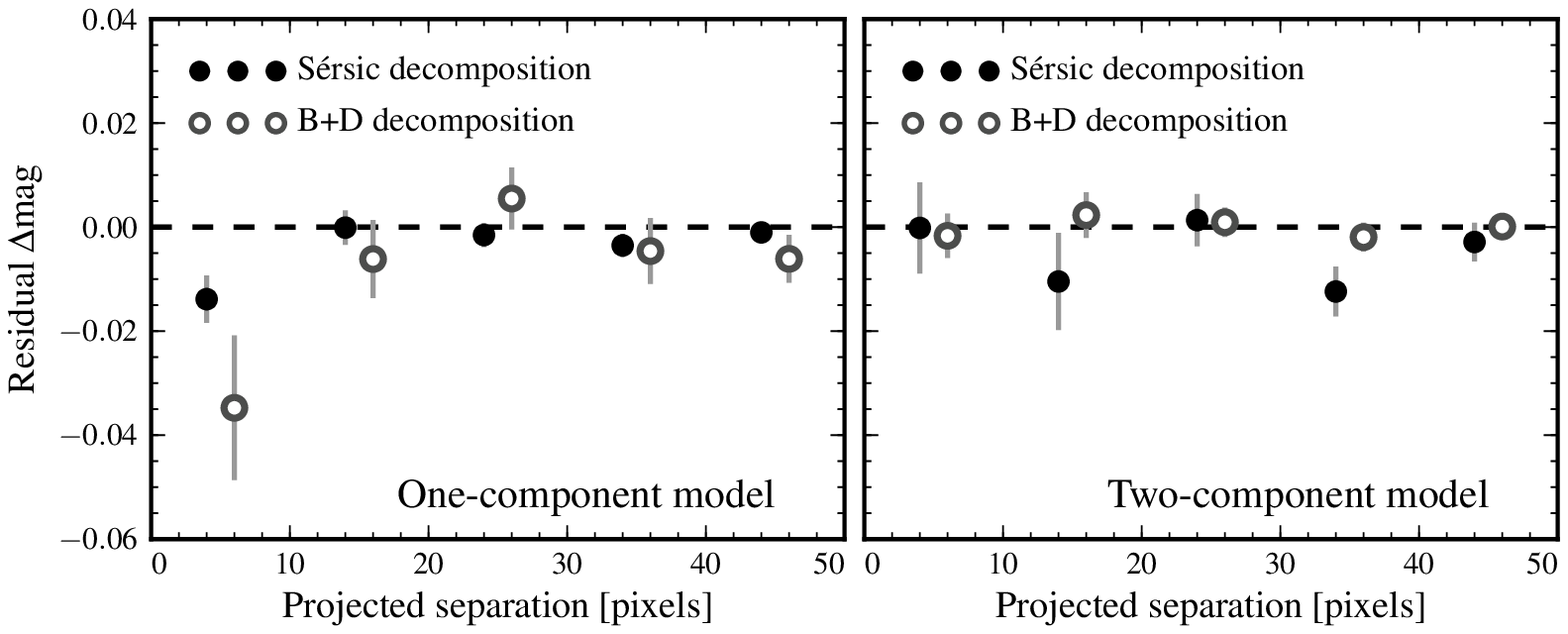}
\caption{Recovered $r$-band flux as a function projected separation to the nearest neighbor.  Filled and open circles show the offset in recovered $r$-band flux when galaxy models are fitted with either a S\'ersic or bulge+disk profile, respectively, where the width reflects the standard error on the mean.  In each case, the residual flux is computed relative to isolated galaxies with a similar apparent magnitude and structure to account for underlying systematics effects.  Left and right panels show results for single- and two-component simulated galaxies.}
\label{fig:crowding}
\end{figure}

As shown in Figures \ref{fig:total_flux_bd} and \ref{fig:total_flux_sersic}, the accuracy of {\sc gim2d}'s flux measurement varies with both apparent magnitude and structure, and we need to account for these variations in order to assess any residual dependence on near neighbors.  Therefore, for each galaxy we compute the flux offset relative to isolated models---i.e. those with no near neighbors---in a moving window of $\Delta m_r$ = 0.25 mag and $\Delta$\btr = 0.2 or $\Delta n$=2 (depending on the underlying model) centered on the galaxy.  This ``residual'' offset can then be used to assess the effects of crowding \emph{independent} of the global trends shown in Figures \ref{fig:total_flux_bd} and \ref{fig:total_flux_sersic}.  In Figure \ref{fig:crowding} we plot the residual offset measured in this way as a function of projected separation in pixels for both one- and two-component simulated galaxies.  The {\sc gim2d} fits are extremely well behaved at separations greater than 10 pixels ($\sim4^{\prime\prime}$), and any residual dependence of the recovered flux on nearby neighbors is typically $<0.02$~mag, i.e. less than the typical photometric uncertainty quoted in Section \ref{flux_recovery}.  Even in the worse case scenario of fitting a (simulated) single-component model with a bulge+disk profile at small separations, the $\sim$0.04 mag offset corresponds to $<$0.02 dex in terms of the derived mass.  While we show results only for the $r$ band, we have verified that any trends with separation are mirrored by the $g$-band models, and show no evidence for a dependence of galaxy color on the presence of near neighbors. 

\end{appendix}

\end{document}